\documentclass[a4paper,fleqn]{cas-sc}

\usepackage[authoryear]{natbib}
\def\tsc#1{\csdef{#1}{\textsc{\lowercase{#1}}\xspace}}
\tsc{WGM}
\tsc{QE}
\tsc{EP}
\tsc{PMS}
\tsc{BEC}
\tsc{DE}
\usepackage{amsmath}
\usepackage{subcaption}
\newtheorem{definition}{Definition}%
\newtheorem{theorem}{Theorem}%
\newtheorem{corollary}{Corollary}

\usepackage{bbm}
\usepackage{soul} % For strikeout

\usepackage{xcolor}
\usepackage[normalem]{ulem}
\begin{document}
\let\WriteBookmarks\relax
\def\floatpagepagefraction{1}
\def\textpagefraction{.001}

\shorttitle{Modelling Evolutionary Power Spectral Density Functions of Strong Earthquakes Via Copulas}

\shortauthors{Ba\~nales et~al.}

\title[mode = title]{Modelling Evolutionary Power Spectral Density Functions of Strong Earthquakes Via Copulas}                      
%\tnotemark[1,2]

%\tnotetext[1]{This document is the results of the research
%   project funded by the National Science Foundation.}

%\tnotetext[2]{The second title footnote which is a longer text matter
%   to fill through the whole text width and overflow into
%   another line in the footnotes area of the first page.}

\author[1]{Isa\'ias Ba\~nales}[%type=editor,
                        %auid=000,bioid=1,
                        %prefix=M.Sc.,
                        % role=Ph.D. Student,
                        orcid=0000-0003-3685-8124]

\cormark[1]

\fnmark[1]

\ead{isaias@ciencias.unam.mx}

%\ead[url]{www.cvr.cc, cvr@sayahna.org}

%\credit{Conceptualization of this study, Methodology, Software}

\affiliation[1]{organization={Centro de Investigaci\'on en Matem\'aticas},
    addressline={Jalisco S/N}, 
    city={Guanajuato},
    % citysep={}, % Uncomment if no comma needed between city and postcode
    postcode={36023}, 
    country={Mexico}}

\author[1]{J. Andr\'es Christen}[orcid=0000-0002-5795-4345]
\fnmark[2]
\ead{jac@cimat.mx}
\author[2]{Josu\'e Tago}[]

\fnmark[3]
\ead{josue.tago@gmail.com}
%\ead[URL]{www.sayahna.org}

%\credit{Data curation, Writing - Original draft preparation}

\affiliation[2]{organization={Universidad Nacional Aut\'onoma de M\'exico},
    city={Mexico City},
    % citysep={}, % Uncomment if no comma needed between city and postcode
    postcode={04510}, 
    %state={Trivandrum},
    country={Mexico}}

\cortext[cor1]{Isa\'ias Ba\~nales}
%\cortext[cor2]{Principal corresponding author}

%\fntext[fn1]{This is the first author footnote. but is common to third
 % author as well.}
%\fntext[fn2]{Another author footnote, this is a very long footnote and
 % it should be a really long footnote. But this footnote is not yet
  %sufficiently long enough to make two lines of footnote text.}

%\nonumnote{This note has no numbers. In this work we demonstrate $a_b$
 % the formation Y\_1 of a new type of polariton on the interface
  %between a cuprous oxide slab and a polystyrene micro-sphere placed
  %on the slab.
  %}

\begin{abstract}
This paper proposes a new approach for analyzing seismic accelerograms using the evolutionary Power Spectral Density function (ePSDF). The accelerogram of an earthquake can be accurately modeled and simulated from its spectrogram, based on the oscillatory stochastic processes theory. To adequately characterize a spectrogram that is consistent with the response spectra, a parametric model with 16 parameters is proposed. This model describes the square of the amplitude spectrum, an envelope of the square of the accelerogram, and a copula that constructs a time-frequency model from the time and frequency marginals.  The use of copulas to model a bivariate probability distribution is a common practice in statistics, particularly when the marginal distributions are known. The periodogram can be viewed as an unnormalized probability density function, where the total energy serves as the normalization constant, since the total energy of a seismic motion is always finite. Additionally, a reduced model consisting of only 10 parameters is presented, which may be especially valuable when only shear wave effects are relevant.
\end{abstract}

% \includegraphics{figs/grabs.pdf}

%\item Low dimensional representation of an accelerogram
%\item Design a framework for attenuation laws that does not rely on Boore's random vibration theory. 
%\item The framework is compatible with previous attenuation laws for amplitude spectrum and accelerogram time envelopes

% Keywords

\maketitle

\section{Introduction}
\label{sec:Introduccion}  

Analyzing earthquakes in the Cocos plate is a significant challenge for Mexican seismology, since strong ground motions may cause severe structural damage in Mexico City and other regions of the country.   Several authors, such as  \cite{arroyo2010strong},  \cite{ordaz1994bayesian} or \cite{castro1993attenuation}, have made important advances in analyzing the amplitude spectra for the southern region of Mexico.

As discussed in \cite{kramer1996geotechnical}, the response spectrum is a crucial ground motion parameter in earthquake engineering.  To assess seismic risk, it is necessary to calculate response spectra directly from available earthquake information.

One of the most influential papers  for simulating accelerograms compatible with response spectrum is \cite{boore2003simulation}; his algorithm,  discussed in his section \textit{Obtaining ground motions},  works on Fourier domain after applying a filter based on the envelope over time of the seismic acceleration.  In that paper a random vibration theory is proposed based on the  main  formula of \cite{cartwright1956statistical} which requires a stationary process, an assumption that, evidently,  earthquake accelerograms do not fulfill. 

Additionally, certain decisions made in \cite{boore2003simulation}, such as the filtering and windowing of the white noise, appear to lack clear justification. These factors, combined with the assumption of stationarity, were the primary motivations behind this study. We present a theoretical approach based on oscillatory process \citep{priestley1965evolutionary};  the new approach allows us to work with a theoretically well-based envelope of the earthquake over the time,  in contrast to the modulation in  \cite{boore2003simulation} based on
"\textit{I find a good comparison between response spectra computed using the box and exponential windows}".

 To avoid assuming a second-order stationary process, the oscillatory stochastic processes theory is used as proposed by \cite{priestley1965evolutionary}.  This theory facilitates the utilization of spectrograms as effective estimators for the Evolutionary Power Spectral Density Function (ePSDF) of an underlying stochastic process responsible for generating the accelerogram.  Moreover, the spectrogram is used in this work to  analyze the energy released in time and frequency separately to finally connect them using the probabilistic theory of copulas.

In the context of signal analysis, copulas have been used to model simulated spectrograms of chirps, as presented in \cite{boashash2015time} and \cite{davy2003copulas}. 

A few novel papers have introduced the use of oscillatory stochastic processes in seismology. In \cite{vlachos2016multi} a mixture of Kanai-Tajimi filters with time-dependent parameters is proposed, while in \cite{broccardo2017spectral} a filter with frequency dependence is used.

In \cite{vlachos2016multi} the idea is to sum multiple Kanai-Tajimi filters with time-variant parameters.  They  present the next formula
\begin{equation}
\label{Eq:Valchos}
S(\mathfrak{f}, t)=|\text{HP}(\mathfrak{f})| \sum_{k=1}^2 S_0^k (t) \frac{1+ \Big( 2\xi^k(t)\mathfrak{f}/ \mathfrak{f}^k_g(t)\Big)^2}{\Big(1-\Big(\mathfrak{f}/ \mathfrak{f}^k_g(t)\Big)^2 \Big)^2+ \Big(2\xi^k(t) \mathfrak{f}/ \mathfrak{f}^k_g(t)\Big)^2}
\end{equation}
where $\{\xi^k(t), \mathfrak{f}^k_g(t),S_0^k (t) \}$ are a time varying modal dominant frequency, modal apparent damping ratio,  modal participation factor with respect to the $k$ mode,  and $\text{HP}$ is a high pass filter.

The necessity of imposing a high pass filter in \eqref{Eq:Valchos} arises from the fact that Kanai-Tajimi filters do not have the value 0 at frequency 0.  Also,  the modelling and estimation of $\{\xi^k(t), \mathfrak{f}^k_g(t),S_0^k (t) \}$ produce a non interpretable model that does not allow to incorporate previous models for the amplitude frequency as  the $\omega^2$ model for amplitude spectrum \citep{brune1970tectonic} or models for the accelerogram envelope as \cite{stafford2009energy}.

In the context of attenuation laws using ePSDF only the paper \cite{vlachos2018predictive} is available.  Its model proposes a 180 parameter regression.  Also its parameters are not interpretable,  and they do not quantify the uncertainty of their estimations. 

The framework presented in this paper aims to simplify the attenuation law for response spectra by first modeling the amplitude spectrum, the envelope of the accelerogram over time, a copula, and the total energy of the accelerogram. This approach results in an interpretable model that is informed by physical considerations.

The seismic data were provided by the accelerographic network provided by the Engineering Institute of UNAM (RAII UNAM,  by its acronym in Spanish), product of the instrumentation and processing work of the Seismic Instrumentation Unit. The data is distributed through the Accelerographic Database System on the web: \url{https://aplicaciones.iingen.unam.mx/AcelerogramasRSM/ }.

The data used in section \ref{sec:Examples} is a subset of the records curated in \cite{garcia2009influence} and used in \cite{arroyo2010strong}. The ground motions are interplate thrust fault events with $M_w\geq 5$ and epicentral distance, $R$, between 100 and 500 km. For those records that did not contain $M_w$, they were collected from the database of Global Centroid-Moment-Tensor (CMT) Project (more information about the CMT project can be consulted in \cite{ekstrom2012global}). 

The outline of the paper is as follows.  In section~\ref{sec:background} the essential theory for understanding the model is presented, which includes elementary results in stochastic oscillatory processes and copulas and why they are useful to model accelerograms. Section \ref{sec:Model} presents a parametric way to describe an accelerogram, as well as the physically informed constraints.  Examples of the model, a discussion of the errors, and a reduced version that works only with the $S-$wave are presented in section \ref{sec:Examples}. Section \ref{sec:Atenuaciones} talks briefly about further work to incorporate attenuation laws into the framework presented in this work. Finally section \ref{sec:Conclusion} summarizes the results obtained in this work.

%%%%%%%%%%%%%%%%%%%%%%%%%%%%%%%%%%%%%%%%%%%%%%%%%

%%%%%%%%%%%%%%%%%%%%%%%%%%%%%%%%%%%%%%%%%%%%%%%%%

\section{Theoretical background}\label{sec:background}

In this section, a brief review of the essential concepts for understanding the model in section \ref{sec:Model} is presented. The concept of oscillatory stochastic processes is introduced, as well as a statistical method for estimating the corresponding ePSDF. The ePSDF is a generalization of the classical spectral density using Fourier analysis for non-stationary oscillatory processes.

\label{sec:Theoretical}

\subsection{Basic Theory on Oscillatory Stochastic Processes}
\label{sec:BasicTheory}

In \cite{priestley1965evolutionary} the concept of \textit{oscillatory processes} is introduced for the first time, which extends the family of stochastic processes admitting a spectral representation.

If there is a complex stochastic process $\{X(t): t\in \mathbb{R}\}$,  which is trend free,  i.e. $\mathbb{E}[X(t)]=0$ for all $t$, and if its covariance function, defined by 
$$
R_{s,t}=\mathbb{E}[X(s)X^*(t)],
$$
may be written as 
\begin{equation}
R_{s,t}=\int_\mathbb{R} \phi_s(\omega)\phi_t^*(\omega)d\mu(\omega),
\label{eq:corr}
\end{equation}
where $*$ denotes the complex conjugate, for functions $\{\phi_t(\omega): t\in \mathbb{R}\}$ and a measure $\mu(\omega)$ defined on the real line. Then, as it is presented in \cite{priestley1965evolutionary}, the process admits the representation
\begin{equation}
X(t)=\int_\mathbb{R}\phi_t(\omega)dZ(\omega),
\label{eq:represt}
\end{equation}
where $Z(\omega)$ is a self orthogonal process,  i.e.
$$
Cov(Z(\omega),Z(\lambda))=0 \quad\text{for all $\omega\neq \lambda$}
$$
and 
$$
\mathbb{E}[(dZ(\omega))^2]=d\mu(\omega).
$$
In addition, in order to have $\text{Var}(X_t)<\infty$ for all $t$, it is required that $\phi_t$ is quadratically integrable.  Over the paper it will be assumed that $\text{Var}(X_t)<\infty$.

If $\phi_t(\omega)=e^{i\omega t}$, the classical spectral representation of an homogeneous stochastic process is retrieved.   Further details about spectral representations for stationary processes can be consulted in \cite{wei2006time}.

In \cite{priestley1965evolutionary} the next definition is presented.

\begin{definition}
For a $t$ fixed, $\phi_t(\omega)$
 will be said to be an oscillatory function if for some  (necessarily unique) $\theta(\omega)$, it may be written in the form 
$$
\phi_t(\omega)=A_t(\omega) e^{i\theta(\omega)t},
$$
where $A_t(\omega)$  is of the form
$$
A_t(\omega)=\int_{\mathbb{R}} e^{it\theta}dH_\omega(\theta),
$$ 
with $H_\omega(\theta)$ a measure that $\mid dH_\omega(\theta)\mid$ has an absolute maximum at $\theta=0$.
\end{definition}

If $\theta(\omega)$ is a single-valued function of $\omega$ then after a variable change from $\omega$ to $\theta(\omega)$ in \eqref{eq:represt} and redefining $A_t(w)$ and the measure $\mu(w)$, the process can be rewritten as 
$$
X(t)=\int_{\mathbb{R}} A_t(\omega)e^{i\omega t}dZ(\omega).
$$

It is important to remark that $A_t(\omega)$ is a deterministic time-frequency dependent  function (as opposed to classic Fourier), which provides  a more flexible framework to model non-stationary processes.

\begin{definition}
If there exists a family of oscillatory functions $\{\phi_t(\omega)\}$,  in terms of which the process $\{X(t)\}$ has a representation of the form \eqref{eq:represt}, $\{X_t\}$ will be termed an oscillatory process.
\end{definition}

In \cite{priestley1965evolutionary} it is explained that if $0\leq t\leq T$ for $T$ in $\mathbb{R}$,  the representation \eqref{eq:represt} always exists.  Considering this result, the accelerograms can be assumed to be oscillatory processes. It is important to note that this is only a sufficient condition, the theory of oscillatory processes does not include only processes in a compact interval over the time.

In the case of an oscillatory process $\{X_t\}$ its variance can be written as
$$ 
\textit{Var}(X(t))=R_{t,t}=\int_{\mathbb{R}} \mid A_t(\omega) \mid^2d\mu(\omega),
$$
which can be seen as the analogous version of the representation in the stationary case. The evolutionary power spectra $dF_t(\omega)$ is defined  as 

$$
dF_t(\omega)=\mid A_t(\omega)\mid ^2d\mu(\omega),
$$
and if $\mu(\omega)$ is absolutely continuous respect to Lebesgue measure, then 
$$
dF_t(\omega)=f_t(\omega)d\omega,
$$
hence $f_t(\omega)$ will be referred to as the ePSDF.

\subsection{Short Time Fourier Transform to Estimate the ePSDF}

Since the representation \eqref{eq:represt} relates the processes $X(t)$ and $\textit{Var}(X(t))$, the ePSDF can be used to simulate realizations of $X(t)$. 

A way to estimate the ePSDF is using the short time Fourier transform (STFT). According to \cite{allen1977short}, the STFT is

\begin{equation}
\text{STFT}(t,\omega)=\int w(t-\tau)X(\tau) e^{-i\omega\tau}d\tau,
\label{eq:STFT}
\end{equation}
where $w$ is an appropriate window function.  To guarantee that STFT preserve the results of Theorem 7.2 in Priestley (1965), it is necessary that $w$ satisfies
\begin{equation}
\label{eq:ConditionFilter}
\int_\mathbb{R} w^2(t)dt=1,
\end{equation}
which guarantee that the estimator \eqref{eq:Estimator} is asymptotically almost unbiased and also ensuring that the total energy of the signal does not change after estimation. In this work $w(t)$ is of the form 

\begin{equation}
\label{eq:Filter}
w(t)\propto e^{-\frac{t^2}{2 (2k)^2}} \mathbbm{1}(\lvert t \rvert <k)  ,     
\end{equation}
for a $k$ in $\mathbb{R}^+$ previously defined, i.e. a Gaussian window and
$$
\mathbbm{1}(\lvert t \rvert <k) =\begin{cases}
1 & \text{if $\lvert t \rvert <k$}\\
0 & \text{if $\lvert t \rvert \geq k$}\\
\end{cases}
$$

Based on \cite{liang2007simulation}, the estimator of the ePSDF using a realization $x_o(t)$ from $X(t)$ will be the square modulus of the STFT of $x(t)$,  i.e.  
\begin{equation}
\label{eq:Estimator}
\overset{\bullet}f_t(\omega)= \Big |  \int w(t-\tau)x_o(\tau) e^{-i\omega\tau}d\tau\Big |^2,
\end{equation}
also is defined 
\begin{equation}
\label{eq:Estimator1}
\widetilde{f}_t(\mathfrak{f})=2\pi \overset{\bullet}f_t(\omega)
\end{equation}
where $\mathfrak{f}$ is expressed in Hertz instead of angular frequency.

To choose $k$ adequately in \eqref{eq:Filter} it is necessary to understand the units of $\widetilde{f}_t(\mathfrak{f})$,   therefore in this section a dimensional analysis is presented

The Parseval's identity could be extended to oscillatory process,  as is presented in \cite{liang2007simulation} as
\begin{equation}
\label{eq:Parseval}
E_T=\int_0^\infty x^2_o(t)dt= \int_0^\infty \int_0^\infty \widetilde{f}_t(\mathfrak{f})dtd\mathfrak{f},
\end{equation}
which ensures the conservation of total energy $E_T$ in our estimate $\widetilde{f}_t(\mathfrak{f})$.

 Since $X^2(t)$ has units in $\frac{cm^2}{s^4}$ (assuming $X(t)$ recorded in $\frac{cm}{s^2}$),  $E_t$ is in $\frac{cm^2}{s^3}$. Given this consideration, we focus our interest in the units of $\widetilde{f}_t(\omega)$ and its marginals. In \eqref{eq:ConditionFilter}  it is  required that $h^2(t)$ will be squared integrable and its integral will be 1,  so  $w(t)$ must be in $\frac{1}{\sqrt{s}}$,  and by  equation \eqref{eq:STFT} it is easy to observe that $ \widetilde{f}_t(\mathfrak{f})$ is in $\frac{cm}{s^3}$,  as expected from \eqref{eq:Parseval}. Having the units of $ \tilde{f}_t(\mathfrak{f})$, it can be easily asserted that the units of its marginals are $\frac{cm^2}{s^4}$ and $\frac{cm^2}{s^2}$ respectively,  where the marginals are defined as
\begin{align*}
\tilde{g}_1(t)&= \int \tilde{f}_t(\mathfrak{f})d\mathfrak{f},\\
\tilde{g}_2(\mathfrak{f})&= \int \tilde{f}_t(\mathfrak{f})dt.
\end{align*}
where $\tilde{g}_1$ and $\tilde{g}_2$ denotes the marginals in frequency and time respectively.

From equation \eqref{eq:Estimator} and considering an observed accelerogram, $x_0$, in the time interval $(0,b)$, the value of $\tau$ where the integrated function is non-zero is in $\Big(\max(\tau-k,0), \min(\tau+k,b)\Big)$, it can be observed that when $k$ goes to infinite the expression in \eqref{eq:Filter} turns into the constant

$$w(t) =\frac{1}{\sqrt{b}}$$
which replaced in \eqref{eq:Estimator1} results in

\begin{equation}
\label{eq:Marg1}
\int_0^b \tilde{f}_t(\mathfrak{f})dt= \int_0^b \Big | \int_{\mathbb{R}} \frac{1}{\sqrt{b}} x_o(\tau) e^{-i 2\pi\mathfrak{f}\tau}d\tau \Big |^2 dt= A^2(\mathfrak{f}).
\end{equation}

For the marginal in time, when $k$ takes the value 0 it can be computed as

\begin{equation}
\label{eq:Marg2}
\int_{\mathbb{R}^+} \tilde{f}_t(\mathfrak{f})d\mathfrak{f}= \int_{\mathbb{R}^+} \Big | \int_{\mathbb{R}} \delta_t(\tau) x_o(\tau) e^{-i 2\pi\mathfrak{f}\tau}d\tau \Big |^2 dt= x_o^2(t),
\end{equation}
where $\delta_t(\cdot)$ is a Dirac delta function in $t$.

In \cite{allen1977short} it is discussed the requirement of a double application of the Nyquist theorem to determine the number of times and frequencies needed to represent the STFT. The ePSDF presents an uncertainty principle problem, resulting in a trade-off between time and frequency resolution. Higher time resolutions lead to poorer frequency resolutions and vice versa. 

Taking into consideration the uncertainty principle, and equations \eqref{eq:Marg1} and \eqref{eq:Marg2} a good option to choose the optimal time window, $k^*$, is 
\begin{equation}
\label{eq:Koptima}
k^*=\underset{k \text{ in } \mathbb{R}^+ }{\mathrm{argmin}} \left\{ \| \lvert x_o \rvert- \sqrt{\tilde{g}_1}\|_2^2 + \| A- \sqrt{\tilde{g}_2}\|_2^2 \right\},    
\end{equation}

where $\| \cdot \|_2^2$ is the squared of the $L_2$ norm.

According to Parseval's theorem, each term in equation \eqref{eq:Koptima}  represents the energy loss of the signal generated by choice $k$ in both time and frequency domains.

\begin{figure}[!ht]
\centering
\begin{subfigure}{0.45\textwidth}
    \includegraphics[width=\textwidth]{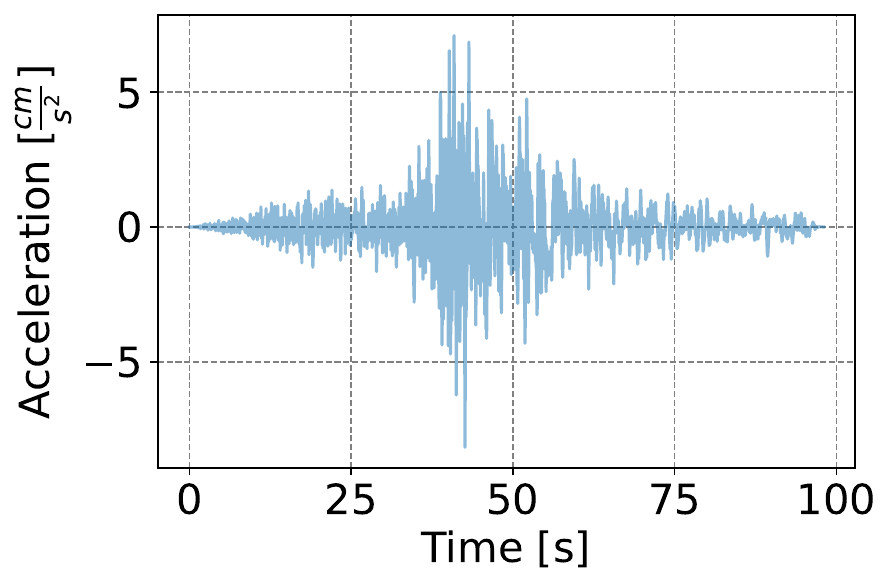}
   % \caption{Record UNIO0301.221.}
    %\label{fig:first}
\end{subfigure}
%\hspace{-.1cm}
\begin{subfigure}{0.45\textwidth}
    \includegraphics[width=\textwidth]{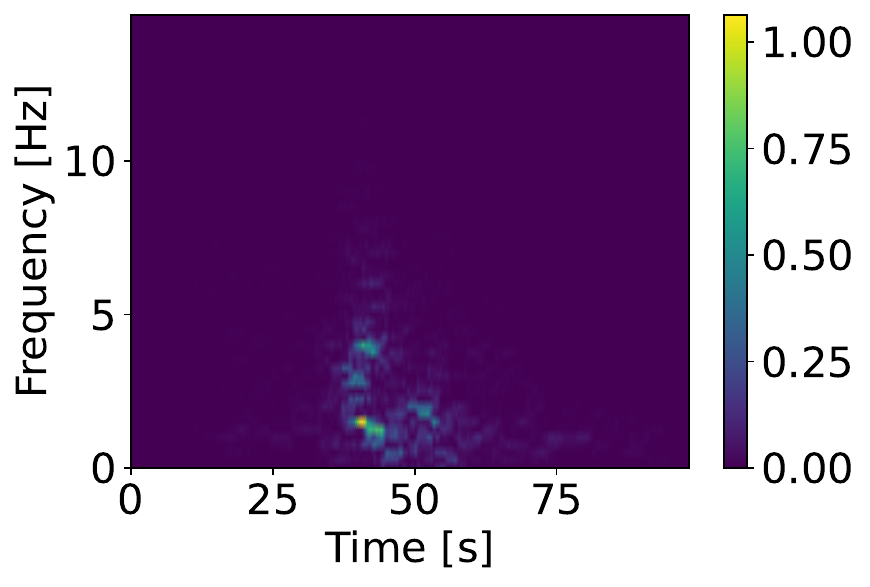}
    %\caption{ePSDF.}
    %\label{fig:second}
\end{subfigure}
%\hspace{0.1cm}
\caption{   Left: Accelerogram of the record. Right: relevant part of its respective ePSDF as estimation of spectral density in Hz, $\widetilde{f}_t(\mathfrak{f})$, where most of its density is concentrated. For the record UNIO0301.221,  of $M_w=7.5$ of 22 January 2003, recorded at La Uni\'on,  state of Guerrero, Mexico, with an epicentral distance of $R=263.59km$ using the North-South chanel.}
\label{fig:AcelSTFT}
\end{figure}

Following this approach and after experimenting with different earthquakes, a time window of 4 seconds was chosen. To increase temporal precision, overlapping intervals of $90\%$ were used.  \cite{vlachos2016multi} used time windows of 3 seconds which produced similar results for the data used in this work. An example of an ePSDF estimated from the STFT  can be seen in Figure \ref{fig:AcelSTFT}.

\begin{figure}[!ht]
\centering
\begin{subfigure}{0.45\textwidth}
    \includegraphics[width=\textwidth]{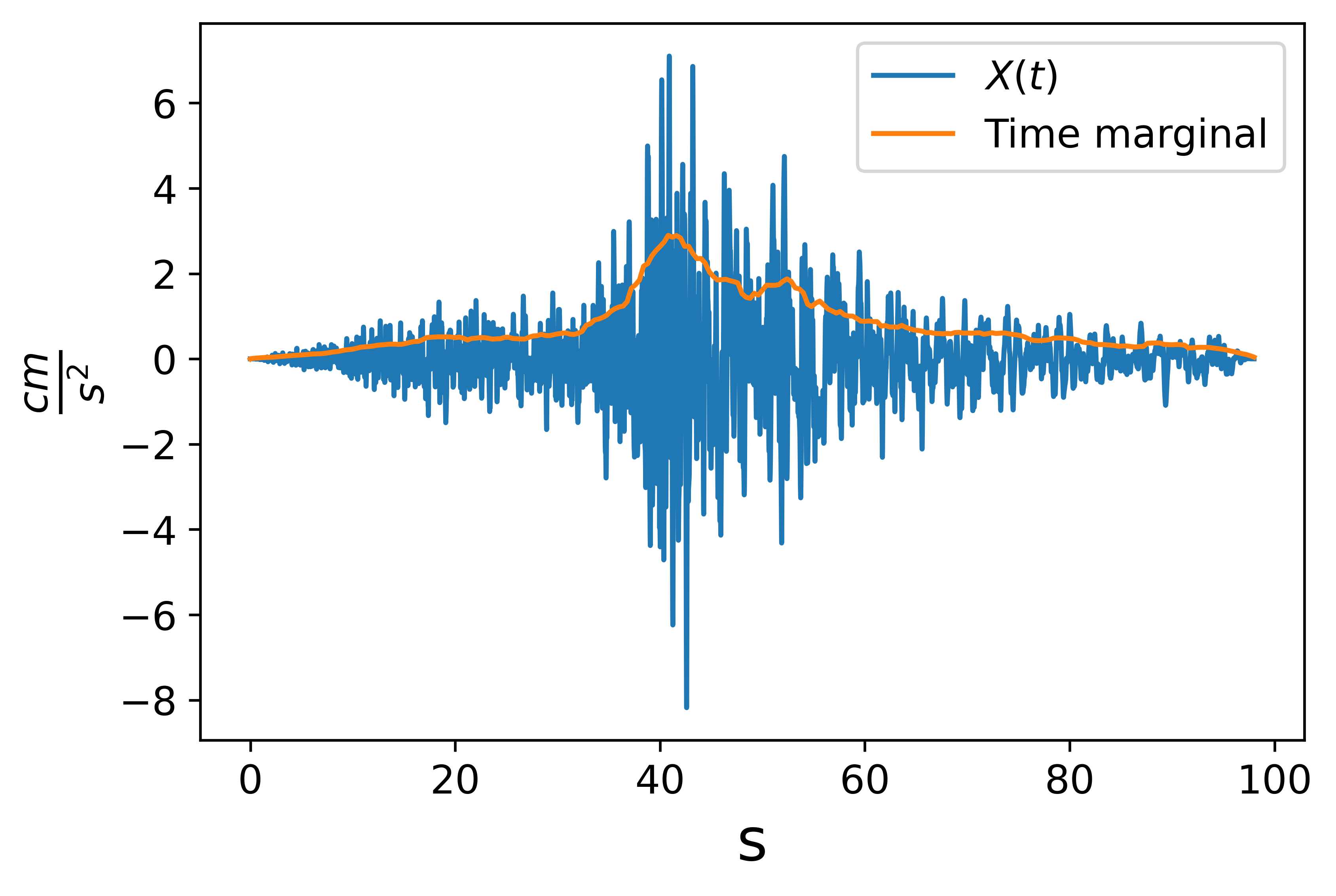}
   % \caption{Record UNIO0301.221.}
    %\label{fig:first}
\end{subfigure}
%\hspace{-.1cm}
\begin{subfigure}{0.45\textwidth}
    \includegraphics[width=\textwidth]{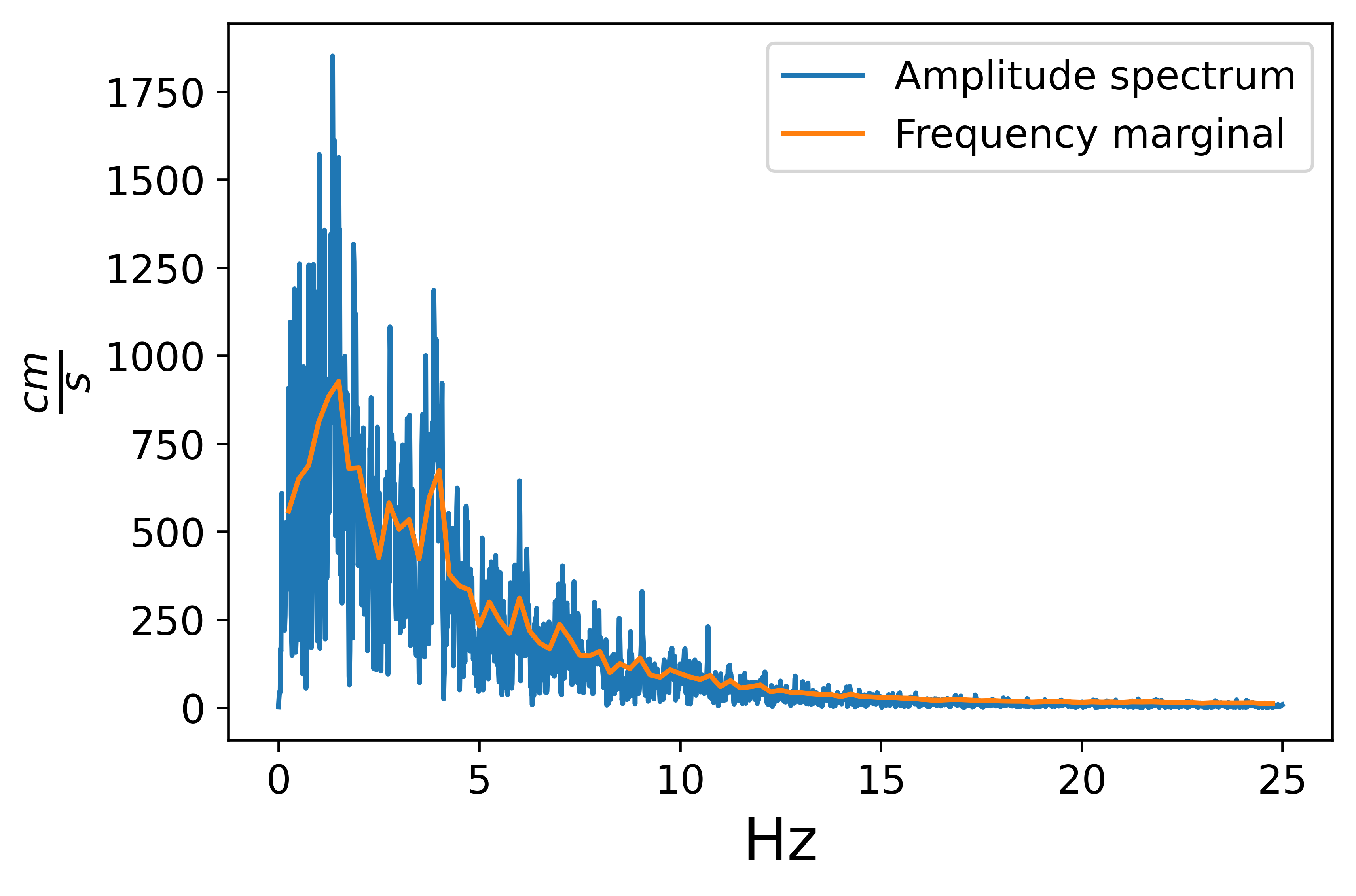}
    %\caption{ePSDF.}
    %\label{fig:second}
\end{subfigure}

\caption{Left: $\sqrt{\tilde{g}_1(t)}$ over  $x_o(t)$. Right: $\sqrt{\tilde{g}_2(\mathfrak{f})}$ over  $AS$. For the record UNIO0301.221}
\label{fig:Envelope}
\end{figure}

It should be noted that the marginal on time,  $\tilde{g}_1(t)$,  has squared units with respect to the acceleration.  On the other hand, the marginal on frequency has squared units with respect to the amplitude spectrum ($A$). For $k^*$ it can be seen in Figure \ref{fig:Envelope} that $\sqrt{\tilde{g}_1(t)}$ and $\sqrt{\tilde{g}_2(\mathfrak{f})}$ approximate very well the accelerogram and the amplitude spectrum of the record respectively.

It is important to note that there are several approaches to estimating the ePSDF besides the STFT, among them those based on wavelets or on the Hilbert-Huang transform.  In this work,    a comparative evaluation of each possible estimation strategy was unnecessary since \cite{bruns2004fourier} proved that, under very mild assumptions, the three approaches are the same.  Other novel approaches exist, \cite{huang2022multi},  \cite{rosen2009local} or \cite{fraser1994multiple}, but theorems demonstrating superior performance to STFT are still lacking. For a review on this topic, the reader is referred to \cite{abreu2017mse}.

\subsection{Simulation of an Oscillatory Stochastic Process}
\label{sec:SimulationOscillatory}

The proposed model considers the accelerogram of a particular earthquake, at a particular location, as a realization of a non-stationary stochastic process, $X(t)$, the actually recorded \textit{observed} accelerogram, $x_o(t)$, can be considered as a realization of such stochastic process $X(t)$. Moreover, any computer simulation $x(t)$ of the modeled accelerogram $X(t)$ is a different realization of such stochastic process and may not be exactly equal to $x_o(t)$.  Note however, that it is not necessary to exactly recover $x_o(t)$, since the model produces synthetic accelerograms, $x(t)$, that have the same distributional properties properties as $X(t)$.

Conceptually, the approach is to view the recorded accelerograms as a realization of an underlying stochastic process, and the aim is to describe such an underlying stochastic process which, according to the previous section, is fully described by its  ePSDF $f_t(\mathfrak{f})$.

In \cite{liang2007simulation}, an approach to the simulation of oscillatory processes is presented.  Using the representation given in \eqref{eq:represt}, they observed that if $X(t)$ is a real process,  as it is in accelerograms,  the conjugated process,  $X^*(t)$, must satisfy
$$
X^*(t)=X(t)=\int_{\mathbb{R}} A^*_t(\omega)e^{-i\omega t}dZ^*(\omega)=\int_{\mathbb{R}} A^*_t(-\omega)e^{i\omega t}dZ^*(-\omega).
$$
Therefore, $A$ and $dZ$ for all $\omega$ satisfy 
\begin{align*}
A_t^*(\omega)&=A_t^*(-\omega)\\
dZ(\omega)&=dZ^*(-\omega).
\end{align*}
Using this idea, \cite{liang2007simulation} prove that
$$
x_o(t)=\sqrt{2}\sum_{k=0}^\infty \Big[ \overset{\bullet}f_t(\omega) \Delta \omega \Big]^{\frac{1}{2}}cos(\omega_k t+\Phi_k),
$$
where $w_k=k\Delta \omega$, $\Delta \omega\to 0$,  and $\Phi_k$ are random variables, independent and identically distributed as uniforms in $[0,2\pi]$. In practice, it is used a finite sum as
\begin{equation}
\label{eq:SimulaFormula}
x(t)=\sqrt{2}\sum_{n=0}^N \Big[  \overset{\bullet}f_t(\omega) \Delta \omega \Big]^{\frac{1}{2}}cos(\omega_n t+\Phi_n),
\end{equation}

where $\Delta\omega=\frac{\omega_u}{N}$,  $\omega_n=n\Delta \omega$ and $\omega_u$ is the Nyquist frequency.  It is important to remember that the evaluation must be performed at the times and frequencies required by the model; in particular, the STFT has no continuity in time or frequency.

\begin{figure}[!ht]
\centering
\begin{subfigure}{0.45\textwidth}
    \includegraphics[width=\textwidth]{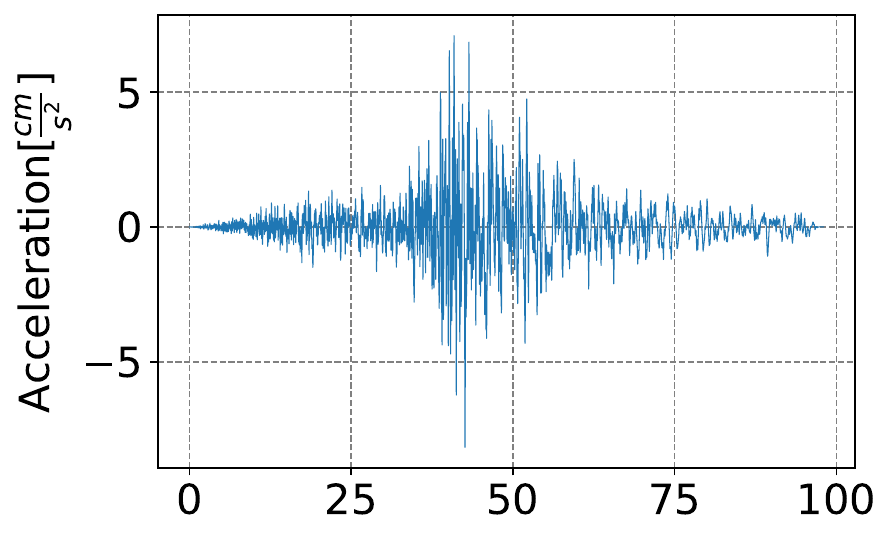}
   % \caption{Record UNIO0301.221.}
    %\label{fig:first}
\end{subfigure}
%\hspace{-.1cm}
\begin{subfigure}{0.45\textwidth}
    \includegraphics[width=\textwidth]{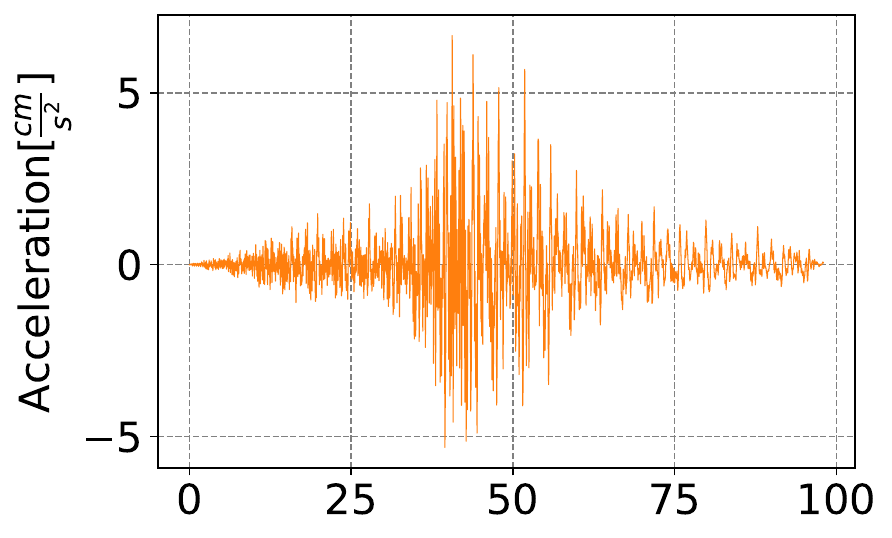}
    %\caption{ePSDF.}
    %\label{fig:second}
\end{subfigure}
%\hspace{0.1cm}

\begin{subfigure}{0.45\textwidth}
    \includegraphics[width=\textwidth]{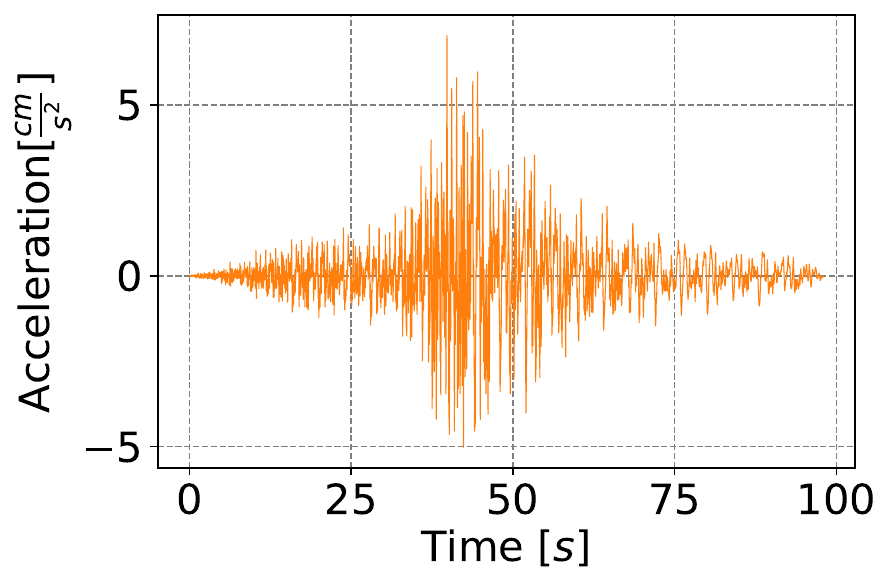}
    %\caption{Frequency Marginal density.}
    %\label{fig:first}
\end{subfigure}
%\hspace{0.1cm}
\begin{subfigure}{0.45\textwidth}
    \includegraphics[width=\textwidth]{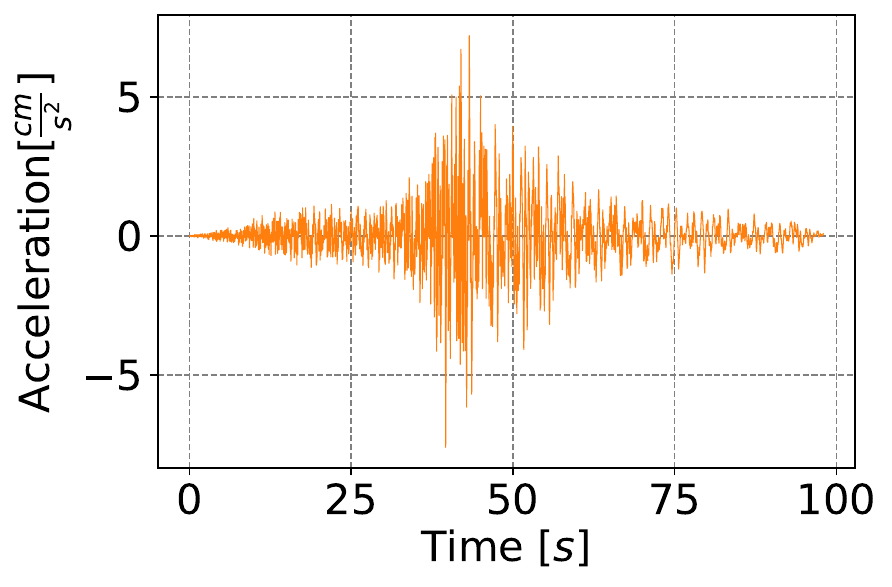}
       % \caption{Time Marginal density.}
    %\label{fig:second}
\end{subfigure}
%\hspace{0.1cm}
\caption{Observed accelerogram for the same record presented in figure~\ref{fig:AcelSTFT} and three simulations using the estimated ePSDF presented in Figure \ref{fig:AcelSTFT} and, simulating using equation \eqref{eq:SimulaFormula}.}
\label{fig:Simulations}

\end{figure}

The Figure \ref{fig:Simulations} shows  an observed accelerogram is presented and three simulated records generated from the equation \eqref{eq:SimulaFormula} using $\widetilde{f}_t(\omega)$ of an observed accelerogram.

From equation \eqref{eq:SimulaFormula} it can be seen that the phase simulation is based solely on the generation of uniform random variables,  unlike  \cite{sato2013fractal} which proposes the use of a fractional Brownian motion to simulate phases.

\subsection{Copulas}
\label{Copulas}
Since $\tilde{f}_t(\mathfrak{f})$ is a non-negative function that preserves the energy of $X(t)$,  which is a finite number as presented in \cite{liang2007simulation},  it  can be seen  as a non-normalized probability density function (p.d.f.). Accordingly, we define the p.d.f. version as 
\begin{equation}
\label{eq:Normalized}
{f}_t(\mathfrak{f})=\frac{1}{Z}\widetilde{f}_t(\mathfrak{f}),
\end{equation}
where 
$$
Z=\int_0^\infty \int_0^\infty \widetilde{f}_t(\mathfrak{f})dt d\mathfrak{f}.
$$

The advantage of working with $f_t(\mathfrak{f})$ as a p.d.f.  is that probability theory  can be utilized,  in particular   the marginals in frequency and time can be computed to recover the joint density using copulas.  For a review on copulas see   \cite{nelsen2007introduction}.  The definition of a copula and Sklar's theorem are also presented in the appendix \ref{sec:BasicsCopulas}.

There are many copulas regularly used in applications \citep{nelsen2007introduction}.  A widely used copula is the Gaussian copula, which is associated with the Gaussian distribution,  and is defined as
\begin{align*}
C_\rho(u,v)&=\mathcal{N}_\rho(\Phi^{-1}(u),\Phi^{-1}(v))\\
&=\frac{1}{2\pi\sqrt{1-\rho^2}}\int_{-\infty}^{\Phi^{-1}(u)}\int_{-\infty}^{\Phi^{-1}(v)}\exp\Big(-\frac{1}{2}{\begin{bmatrix}
s&t
\end{bmatrix}
\begin{bmatrix}
1&\rho\\
\rho&1
\end{bmatrix}^{-1}
{\begin{bmatrix}
s \\
t
\end{bmatrix}}}\Big)dsdt,
\end{align*}
where $\mathcal{N}_\rho$ is the joint cumulative distribution function of a bivariate Gaussian distribution with mean 0,  variances 1,  correlation $\rho$ and  $\Phi^{-1}$ is the quantile function of a standard univariate Gaussian distribution.

Section \ref{sec:TF} uses a Gaussian copula to relate the time and frequency marginal distributions. There are two important reasons for choosing a Gaussian copula.  First $\rho$ controls the correlation of the joint distribution and second $C_\rho$ has a density with the following explicit formulation 
$$
c_\rho(u,v)= \frac{1}{\sqrt{1-\rho^2}}\exp\Big(-\frac{1}{2}{\begin{bmatrix}
\Phi^{-1}(u)&\Phi^{-1}(v)
\end{bmatrix}\Big(
\begin{bmatrix}
1&\rho\\
\rho&1
\end{bmatrix}^{-1}
-\begin{bmatrix}
1&0\\
0&1
\end{bmatrix}
\Big)
{\begin{bmatrix}
\Phi^{-1}(u) \\
\Phi^{-1}(v)
\end{bmatrix}}}\Big).
$$

In  \cite{davy2003copulas} there is a detailed discussion of the use of Copulas for signal analysis. It is important to remark that \cite{davy2003copulas} work with a  complicated  family of copulas with a complicated expression of the density copula,  which causes a difficult physical interpretation. To avoid this, in this work it is proposed the Gaussian copula, which is less flexible,  but seems capable to model accelerograms correctly.

The Gaussian copula is useful for low-dimensional representation of accelerogams because it has only one parameter, $\rho$. Section  \ref{sec:TF} discusses how $\rho$ is restricted to create realistic models for accelerograms. Additionally, the supplementary material presents accelerograms with different conditions to emphasize that the Gaussian copula is flexible enough for our purposes.

Moreover, other copulas were tested as the Clayton and Student's t  copulas \citep{nelsen2007introduction}  as well as those presented in \cite{chesneau2023extensions}, among others.  No significant improvements where obtained with respect to the Gaussian copula, to justify their more complex expressions.

%%%%%%%%%%%%%%%%%%%%%%%%%%%%%%%%%%%%%%%%%%%%%%%%%

%%%%%%%%%%%%%%%%%%%%%%%%%%%%%%%%%%%%%%%%%%%%%%%%%

\section{A model to approximate the normalized ePSDF }
\label{sec:Model}

 This section presents an approach for modeling the marginal densities of a normalized ePSDF on time and frequency.  The strategy involves initially modeling and approximating the marginals of the normalized ePSDF. Then, using a Gaussian copula, an approximation of the full normalized ePSDF is obtained. This,  along with the calculated total energy, provides a distributional description of the underlying stochastic process for the accelerogram.  This approximation will be parameterized with a total of 16 parameters, resulting in a significant reduction in dimensionality compared to the full accelerogram, which has $50-200$ samples per second.

\subsection{Time domain marginal}
\label{sec:TimeDomain}

To model the time marginal,  it is required an envelope of the energy released along time.  In \cite{stafford2009energy},   a model based on a lognormal distribution is presented to envelope a seismic wave.  The lognormal distribution $f_{[\mu, \sigma^2, l]}$ is given by
$$
f_{[\mu, \sigma^2, l]}(x)= {\displaystyle {\frac {1}{(x-l)\sigma {\sqrt {2\pi }}}}\ \exp \left(-{\frac {\left(\ln \left((x-l)\right)-\mu \right)^{2}}{2\sigma ^{2}}}\right)}\mathbbm{1}\{x>l\},
$$
the location parameter has been added to handle the problem of a non standardized pre sample time in the records.

To take the $P$ and $S-$waves into account simultaneously, the time marginal  is model with a mixture of two lognormal distributions.  To ensure that the $S$ wave releases more energy than the $P-$wave,  phyisically informed restrictions were imposed as is discussed in \ref{sec:ModelFitting}.  The model for the  time marginal  is  formulated  as
\begin{equation} 
\label{eq:Marginalt}
\widehat{g}_1(t)= \pi_1 f_{[\mu_1, \sigma_1^2, l_1]}(t)  + (1-\pi_1) f_{[\mu_2, \sigma_2^2, l_1+\gamma]}(t).
\end{equation}
The intention of the $\gamma$ parameter is to guarantee that the $S-$wave arrives after the $P-$wave, namely $\gamma > 0$.  In Section \ref{sec:ModelFitting},  it is  discussed how prior information about wave velocities was used to further state admissible values for $\gamma$. It is important to note that $(\mu_1, \sigma_1^2, l_1)$ is related to the $P-$wave and $(\mu_2, \sigma_2^2, l_1+\gamma)$ to the $S-$wave.

A further extension of this work could be the addition of prior information based on partition energy among elastic waves as in  \cite{sanchez2011energy}.

\subsection{Frequency domain}
\label{sec:FreqDomain}

A seminal paper on amplitude spectrum modelling  in a Bayesian framework  in Mexico is \cite{ordaz1994bayesian}, which uses the $\omega^2$ model.  A brief survey of amplitude spectrum models is presented in \cite{boore2003simulation} and a model for finite-source in Mexico can  be consulted in  \cite{singh1989analysis}.  

The  amplitud frequency models presented in \cite{boore2003simulation} could  be well approximated by a lognormal distribution. To model the $P-$wave frequency marginal  a single lognormal distribution is used, while for the $S$ case a mixture of two lognormal distributions is proposed. The physical  reason to do this distinction is that $S-$waves could be overlapped producing a multimodal $S-$wave marginal or because the $S-$wave could transport higher energy on a specific frequency. The models for the frequency marginals are 
\begin{align*}
\widehat{g}_{21}(\mathfrak{f})&\propto f_{[\mu_{\mathfrak{f}_1}, \sigma_{\mathfrak{f}_1}^2, 0]}(\mathfrak{f}) \mathbbm{1}_{\mathfrak{f}\leq \mathfrak{f}_{max}} \\
\widehat{g}_{22}(\mathfrak{f})&\propto \Big( m_2 f_{[\mu_{\mathfrak{f}_{21}}, \sigma_{\mathfrak{f}_{21}}^2, 0]}(\mathfrak{f}) +(1-m_2) f_{\mu_{[\mathfrak{f}_{22}}, \sigma_{\mathfrak{f}_{22}}^2, 0]}(\mathfrak{f})\Big) \mathbbm{1}_{\mathfrak{f}\leq \mathfrak{f}_{max}},\\
\end{align*}
The $\widehat{g}_{21}$ is the  model for the  frequency  marginal  before the $S-$wave arrival time,  i.e.  before $l_1+\gamma$,  and $\widehat{g}_{22}$ after the arrival. Note that the location parameters are fixed at 0 unlike as in the time marginal density.

As  presented in \cite{boore2003simulation}, \cite{arroyo2010strong} and \cite{parajuli2017earthquake},  it was chosen $\mathfrak{f}_{max}=15$Hz as higher frequencies are usually not of interest.

\subsection{Time-Frequency domain model using copulas}
\label{sec:TF}

As explained earlier, with the approximations for the time and frequency marginals, the joint density is approximated using a Gaussian copula.

The proposal is to use a Gaussian copula \textit{with negative correlation} to recover the joint distribution $f_t(\mathfrak{f})$, with the following model, namely 
\begin{align}
\label{eq:model}
\widehat{f}_{t}(\mathfrak{f})&:=\widehat{G}_1(l_1+\gamma) c_\rho(\widehat{G}_{11}(t),\widehat{G}_{21}(\mathfrak{f}))\widehat{g}_{11}(t)\widehat{g}_{21}(\mathfrak{f})\\
\nonumber
&+(1-\widehat{G}_1(l_1+\gamma)) c_\rho(\widehat{G}_{12}(t),\widehat{G}_{22}(\mathfrak{f}))\widehat{g}_{12}(t)\widehat{g}_{22}(\mathfrak{f}),
\end{align}
The densities $g_{1i}$ for  $i$ in $\{1,2\}$  are defined as
\begin{align*}
\widehat{g}_{11}(t)\propto \widehat{g}_1(t)\mathbbm{1}(t<l_1+\gamma) \\
\widehat{g}_{12}(t)\propto \widehat{g}_1(t)\mathbbm{1}(t>l_1+\gamma),
\end{align*}
to be probability density functions. The capital letters are the distribution functions of the corresponding density functions,  for example

$$
\widehat{G}_1(x)=\int_{0}^x \widehat{g}_1(u)du. 
$$

In Section \ref{sec:Examples}, there are examples of the fitting of the proposed model to real seismic motions. 

It is important to note that in order to calculate the response spectra, it is necessary to know the energy of the record,  which is easily calculated from 
\begin{equation} \label{eq:total_energy}
E_T=\int X^2(t)dt.    
\end{equation}

\subsection{Model Fitting}
\label{sec:ModelFitting}
Equation \eqref{eq:model} requires the estimation of  following parameters 
$$
\Theta=\{\mu_1,\sigma^2_1,l_1, \mu_2,\sigma_2^2,\gamma,\pi_1,  \mu_{\mathfrak{f}_1},  \sigma^2_{\mathfrak{f}_1},  \mu_{\mathfrak{f}_{21}},  \sigma^2_{\mathfrak{f}_{21}},   \mu_{\mathfrak{f}_{22}}, \sigma^2_{\mathfrak{f}_{22}},  \pi_2,  \rho\}. 
$$

Since the signals have different orders of magnitude before and after the time $l_1+\gamma$, the fitting of the parameters were accomplished using additive Gaussian errors with different variance in each block to define a likelihood in which the variance of the error is also estimated.

Defining $\mathcal{T}$ and $\mathcal{F}$ as the values in which the time and frequency can be evaluated in the STFT, the negative log-likelihood is given by 

\begin{align}
\label{eq:veros}
-\ell(\Theta)&=\sum_{t \in \mathcal{T}}  \sum_{\mathfrak{f} \in \mathcal{F}} \frac{\Big(\widehat{f}_{t}(\mathfrak{f})-f_{t}(\mathfrak{f})\Big)^2}{2 \sigma_{\epsilon_1}^2}\mathbbm{1}(t<l_1+\gamma) \mathbbm{1}(\mathfrak{f}<\mathfrak{f}_{max})\\ \nonumber
&+\sum_{t \in \mathcal{T}}  \sum_{\mathfrak{f} \in \mathcal{F}}
\frac{\Big(\widehat{f}_{t}(\mathfrak{f})-f_{t}(\mathfrak{f})\Big)^2}{2 \sigma_{\epsilon_2}^2}\mathbbm{1}(t\geq l_1+\gamma) \mathbbm{1}(\mathfrak{f}<\mathfrak{f}_{max})\\ \nonumber
&+  \Big(\sum_{\mathfrak{f} \in \mathcal{F}} \mathbbm{1}(\mathfrak{f}<\mathfrak{f}_{max}) \Big) \Big(\sum_{t \in \mathcal{T}}  \log( \sigma_{ \epsilon_1}^2) \mathbbm{1}(t< l_1+\gamma)\\ \nonumber
&+ \sum_{t \in \mathcal{T}} \log( \sigma_{\epsilon_2}^2)\mathbbm{1}(t\geq l_1+\gamma) \Big) 
\end{align}

 To compute a synthetic accelerogram from the model, it is necessary to know the total energy, $E_t$, which can be computed directly from the acceleration record, $X(t)$ (see equation (\ref{eq:total_energy})).  
 
 It can be observed that if the Gaussian error $\sigma_{\epsilon_i}$ in each section is known, the equation \eqref{eq:veros} produces a weighted least squares fitting.

It is important to note that $\sigma_{\epsilon_1}$ and $\sigma_{\epsilon_2}$ are nuisance parameters, so the model needs only 16 parameters to describe the entire history of an accelerogram, which are 
$$
\{\mu_1,\sigma^2_1,l_1, \mu_2,\sigma_2^2,\gamma,\pi_1,  \mu_{\mathfrak{f}_1},  \sigma^2_{\mathfrak{f}_1},  \mu_{\mathfrak{f}_{21}},  \sigma^2_{\mathfrak{f}_{21}},   \mu_{\mathfrak{f}_{22}}, \sigma^2_{\mathfrak{f}_{22}},  \pi_2,  \rho,  E_T\}. 
$$

To guarantee that $\pi_i$ in $i=1,2$ are a mixture parameters, they are constrained over the interval $(0,1)$.  

In order to have a negative correlation in the Gaussian copula $\rho$ is constrained to be less than 0.  The idea that supports the use of a negative correlation between time and frequency is that the Earth is an anelastic medium that acts as a low-pass filter, resulting in that lower frequencies prevail more time than the higher frequencies.  A detailed discussion regarding energy release in seismic waves can be found in \cite{shearer2009introduction}.

The velocity model of southern Mexico was taken from \cite{castro1993attenuation} with $P-$wave velocity of $6.1\frac{km}{s}$ and $S-$wave velocity of $3.5\frac{km}{s}$.

The arrival times can be roughly estimated using epicentral distance, $R$, and the corresponding wave velocity.  The difference between the estimated arrival times of the $P$ and $S-$waves 
$$
T_b = \frac{R}{3.5}-\frac{R}{6.1}
$$
is used to constrain $\gamma$ in $(0.5T_b,1.5T_b)$, letting the actual arrival time difference (i.e. $\gamma$) to be within 50\% above and below the simple estimate $T_b$.

To avoid unidentifiability and to guarantee that the first wave corresponds to the $P-$wave, it is forced in \eqref{eq:Marginalt}  that the density $f_{\mu_1,\sigma^2_1,l_1}$ reaches its maximum before $f_{\mu_2,\sigma^2_2,l_1+\gamma}$ does. Another constraint is that  the peak of $f_{\mu_1,\sigma^2_1,l_1}$ and $f_{\mu_2,\sigma^2_2,l_1+\gamma}$ must be in the interval $(0.5 T_b, 1.5T_b)$, to guarantee that the first energy arrival occurs before the second. 

Finally,  to avoid too flat densities,  the variances of the two lognormals are bounded by $50^2$.  This bound is arbitrary and can be improved if a model of the duration of seismic waves  is used.  Since the $P-$wave register contains  less energy, the last condition $0<\sigma_{\varepsilon_1}^2<\sigma_{\varepsilon_2}^2$ is imposed.

Then the optimization problem to obtain $\Theta$ is

\begin{align*}
&\underset{\Theta}{\mathrm{argmin}} -\ell(\Theta)\\
&\text{subject to}\\
&\Theta>0,\\
&\pi_{i} \text{  in } (0,1), \text{ for $i$ in } \{1,2\},\\
&\sigma_{\epsilon_1}<\sigma_{\epsilon_2},\\
&\gamma \text { in } (0.5T_b,1.5 T_b),\\
&\max_t(f_{\mu_1,\sigma^2_1,l_1}(t))<\max_t(f_{\mu_2,\sigma^2_2,l_1+\gamma}(t)),\\
&\underset{t}{\mathrm{argmax}}(f_{\mu_2,\sigma^2_2,l_1+\gamma}(t))-\underset{t}{\mathrm{argmax}}(f_{\mu_1,\sigma^2_1,l_1}(t)) \text { in } (0.5T_b,1.5 T_b) \text{ and }\\
&{\displaystyle [\exp(\sigma_i ^{2})-1]\exp(2\mu_i +\sigma_i ^{2})}<50^2 \text{ for $i$ in } \{1,2\},
\end{align*} 
where $\Theta>0 $ means that each element of vector $\Theta$ is positive.

This is a difficult nonlinear optimization problem with complex constraints. For the examples presented here the t-walk algorithm \citep{christen2010general} was  used to carry out the required optimization .
The t-walk is actually a Markov Chain Monte Carlo (MCMC) algorithm,  but it can easily be used for optimization as well.  It is derivative free, does not require tuning parameters, domain restrictions are straightforward to include and has several implementations, including one in Python.  

%%%%%%%%%%%%%%%%%%%%%%%%%%%%%%%%%%%%%%%%%%%%%%%%%

%%%%%%%%%%%%%%%%%%%%%%%%%%%%%%%%%%%%%%%%%%%%%%%%%

\section{Examples}
\label{sec:Examples}
In this section, the fitting of the model to real records will be shown. Only two examples are presented in detail but several more can be found in the supplementary material. 

\subsection{Full model examples}
\label{sec:FullExamples}

\begin{figure}[!ht]

\centering
\begin{subfigure}{0.45\textwidth}
    \includegraphics[width=\textwidth]{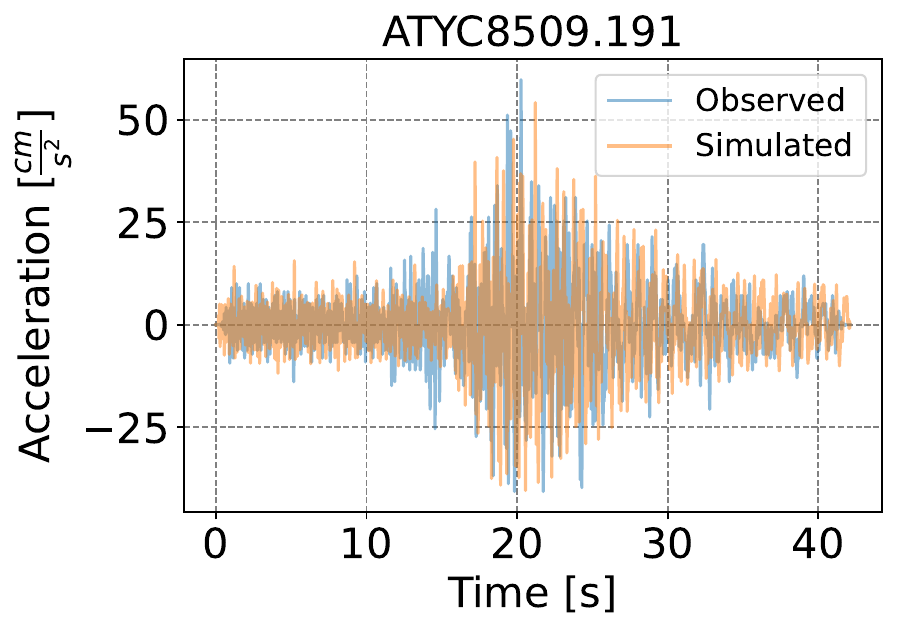}
   % \caption{Record UNIO0301.221.}
    %\label{fig:first}
\end{subfigure}
%\hspace{-.1cm}
\begin{subfigure}{0.45\textwidth}
    \includegraphics[width=\textwidth]{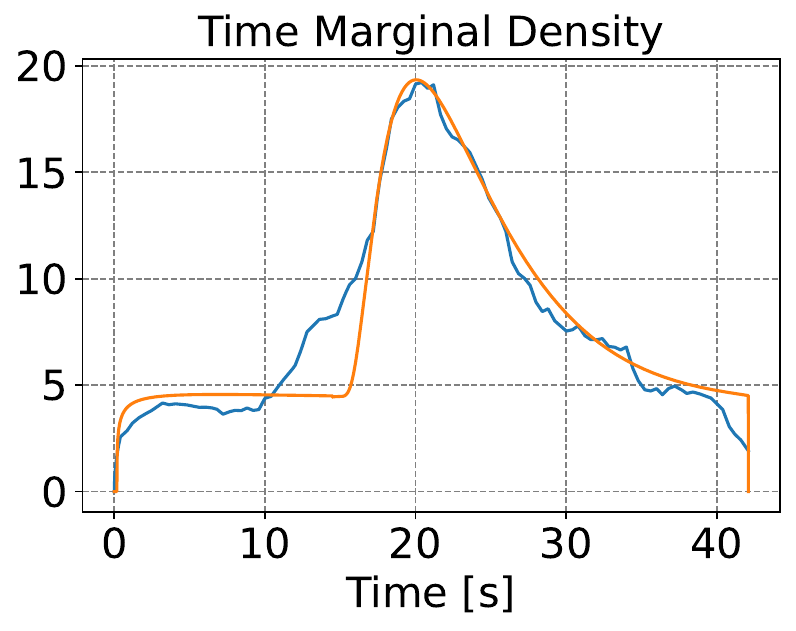}
    %\caption{ePSDF.}
    %\label{fig:second}
\end{subfigure}
%\hspace{0.1cm}

\begin{subfigure}{0.45\textwidth}
    \includegraphics[width=\textwidth]{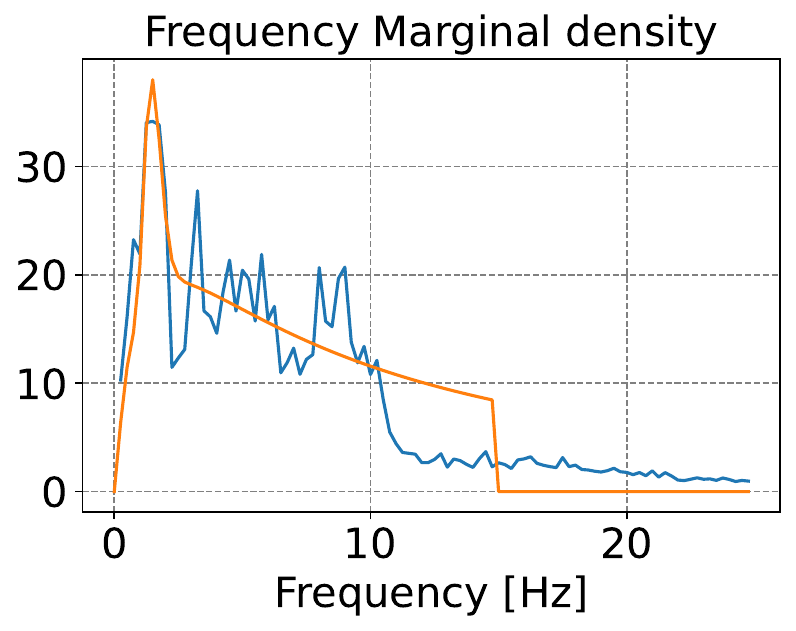}
    %\caption{Frequency Marginal density.}
    %\label{fig:first}
\end{subfigure}
%\hspace{0.1cm}
\begin{subfigure}{0.45\textwidth}
    \includegraphics[width=\textwidth]{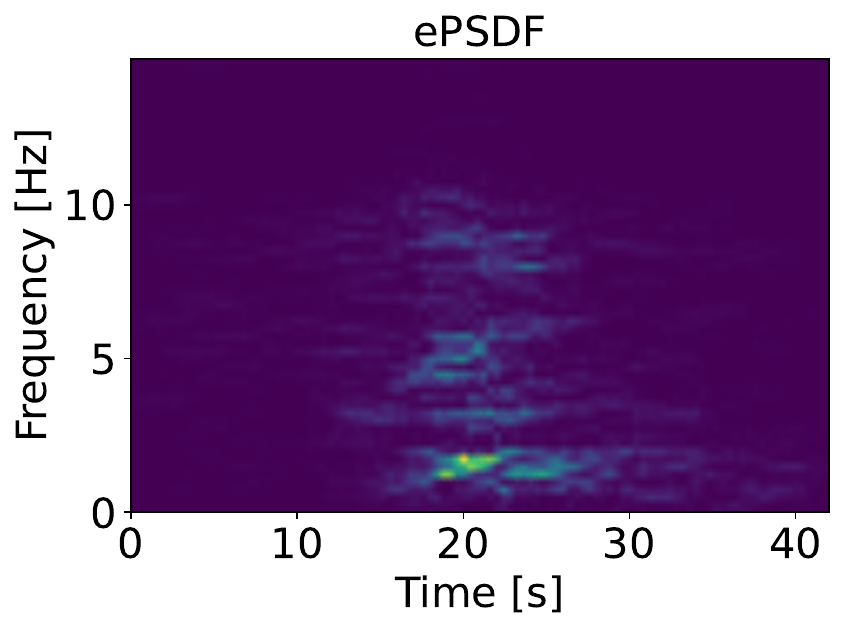}
       % \caption{Time Marginal density.}

    %\label{fig:second}
\end{subfigure}
%\hspace{0.1cm}

\caption{Top left: A simulated accelerogram using our fitted model (orange) plotted over the observed accelerogram (blue). Top right: Time marginal $g_1(t)$ and $\hat{g}_1(t)$. Bottom left: Frequency marginal of $g_{2}(\mathfrak{f})$ and $\hat{g}_{2}(\mathfrak{f})$. Bottom right: $f_{t}(\mathfrak{f})$.  For the record ATYC8509.191,  of $M_w=7.5$ of 19 September 1985, recorded at Atoyac,  state of Guerrero, Mexico, with an epicentral distance of $R=283.09km$ using the North-South chanel.}
\label{fig:ATYC6}
\end{figure}

\begin{figure}[!ht]
\centering
\begin{subfigure}{0.45\textwidth}
    \includegraphics[width=\textwidth]{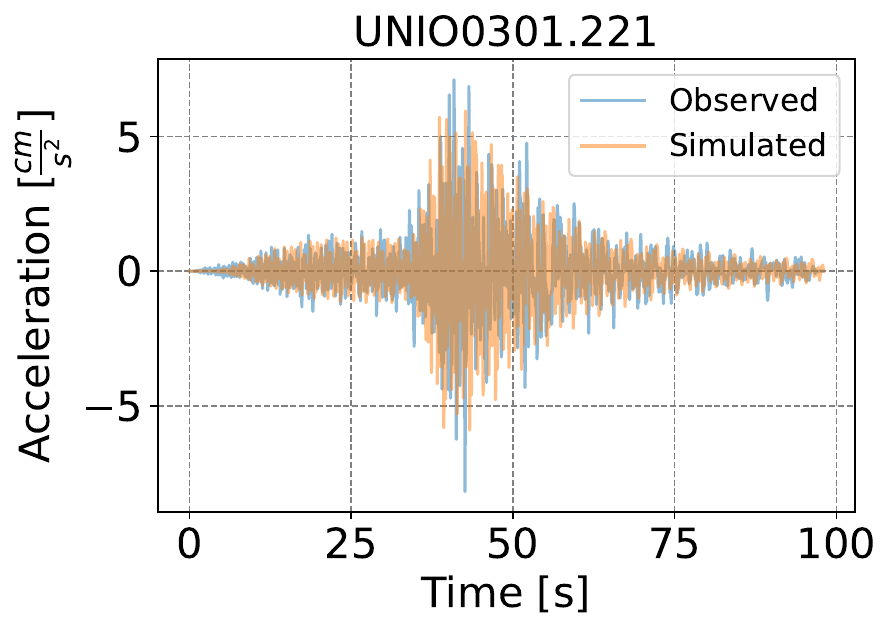}
   % \caption{Record UNIO0301.221.}
    %\label{fig:first}
\end{subfigure}
%\hspace{-.1cm}
\begin{subfigure}{0.45\textwidth}
    \includegraphics[width=\textwidth]{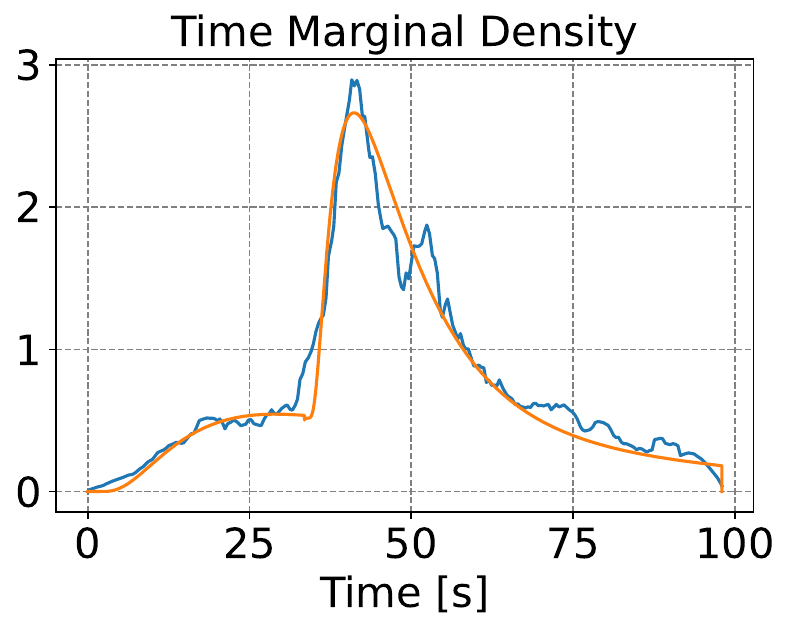}
    %\caption{ePSDF.}
    %\label{fig:second}
\end{subfigure}
%\hspace{0.1cm}

\begin{subfigure}{0.45\textwidth}
    \includegraphics[width=\textwidth]{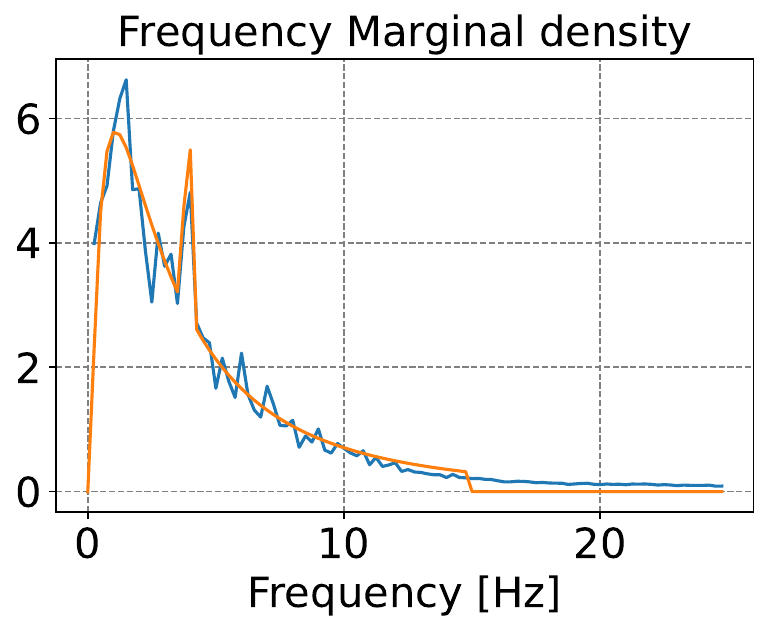}
    %\caption{Fitted model over ePSDF}
    %\label{fig:third}
\end{subfigure}
\begin{subfigure}{0.45\textwidth}
    \includegraphics[width=\textwidth]{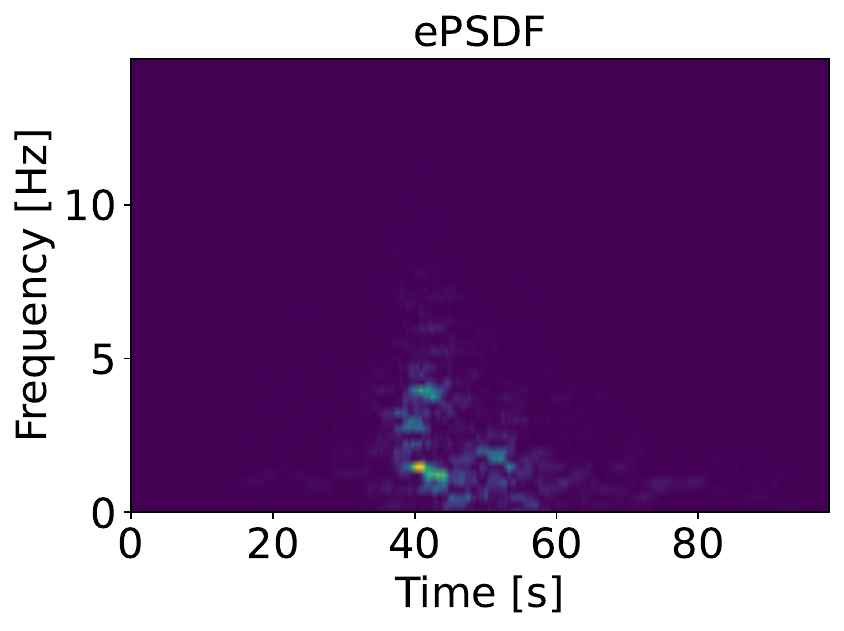}
    %\caption{Frequency Marginal density.}
    %\label{fig:first}
\end{subfigure}
    
\caption{Top left: A simulated accelerogram using our fitted model (orange) plotted over the observed accelerogram (blue). Top right: Time marginal $g_1(t)$ and $\hat{g}_1(t)$. Bottom left: Frequency marginal of $g_{2}(\mathfrak{f})$ and $\hat{g}_{2}(\mathfrak{f})$. Bottom right: $f_{t}(\mathfrak{f})$.  For the record UNIO0301.221}

\label{fig:UNIO6}
\end{figure}

Figures \ref{fig:ATYC6} and \ref{fig:UNIO6} show the fit of the proposed  model to a horizontal component of the records ATYC8509.191 and UNIO0301.221.

The fitted model based on the  copula approach presents satisfactory results for marginal densities on time and frequency,  which means that the energy of our model is correctly released.  Also the response spectra of  the simulated records approximates quite well to the real response spectra, as seen in Figures \ref{fig:ATCY62} and \ref{fig:UNIO62}.

Based on the different examples presented in this section and in the supplementary material,   the proposed  model is flexible and useful for modeling different seismic waves.

One of the major advantages of having a model to simulate accelerograms is that it allows the analysis of how different wave paths change the response spectra at a specific station, even for events with  similar $M_w$ and $R$. The response spectra of the records shown in figures \ref{fig:ATYC6} and \ref{fig:UNIO6} may be seen in figures \ref{fig:ATCY62} and \ref{fig:UNIO62}, respectively.

\begin{figure}[!ht]

\centering

%\hspace{0.1cm}

\begin{subfigure}{0.45\textwidth}
    \includegraphics[width=\textwidth]{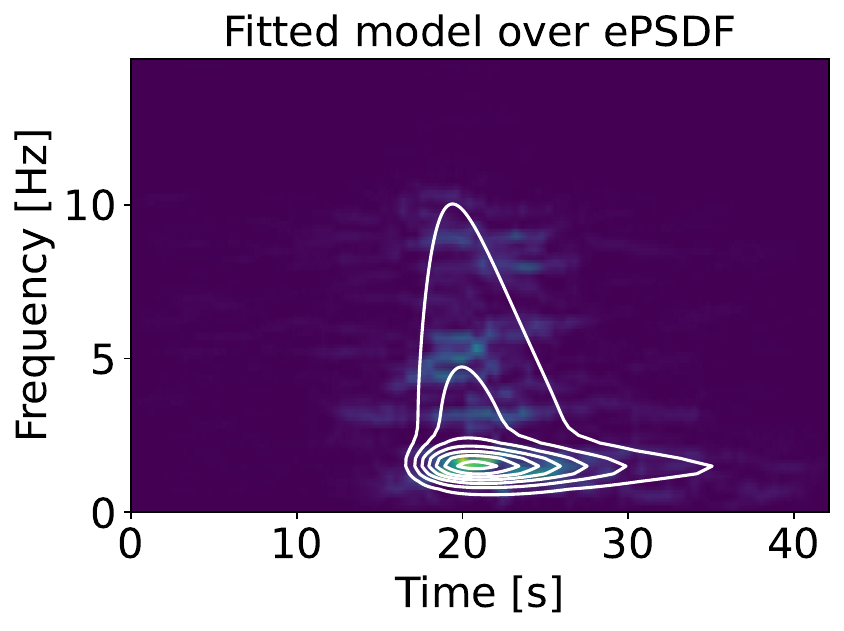}
       % \caption{Time Marginal density.}

    %\label{fig:second}
\end{subfigure}
%\hspace{0.1cm}
\begin{subfigure}{0.45\textwidth}
    \includegraphics[width=\textwidth]{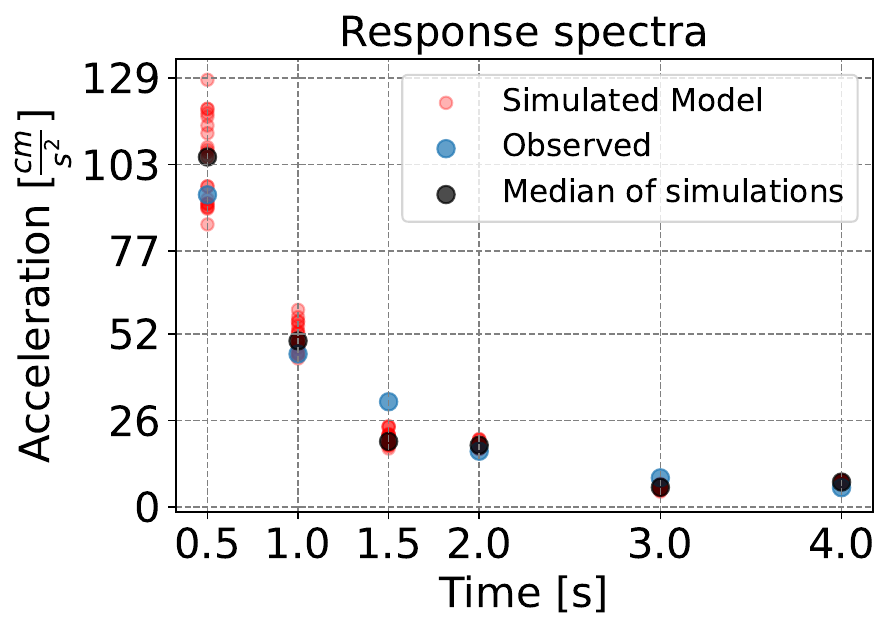}
    %\caption{Response spectra.}
   % \label{fig:third}
\end{subfigure}

\caption{Same accelerogram as in Figure~\ref{fig:ATYC6}.  Left: level curves of our fitted full model  overlapped to the ePSDF.  Right: the response spectra of 20 simulations using our fitted model, and their median, and the response spectra of the observed accelerogram.}
\label{fig:ATCY62}
\end{figure}

\begin{figure}[!ht]

\centering
\begin{subfigure}{0.45\textwidth}
    \includegraphics[width=\textwidth]{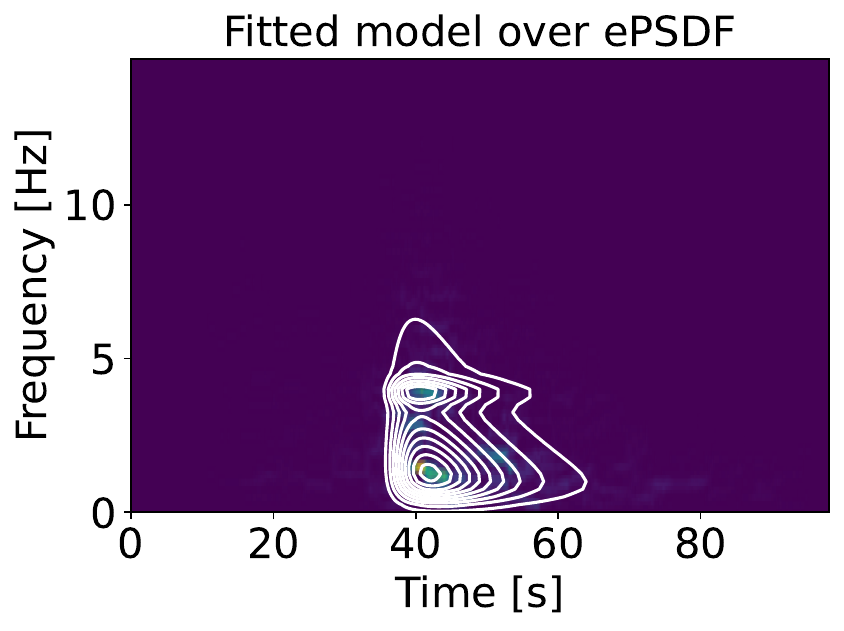}
   % \caption{Record UNIO0301.221.}
    %\label{fig:first}
\end{subfigure}
%\hspace{-.1cm}
%\hspace{0.1cm}
\begin{subfigure}{0.45\textwidth}
    \includegraphics[width=\textwidth]{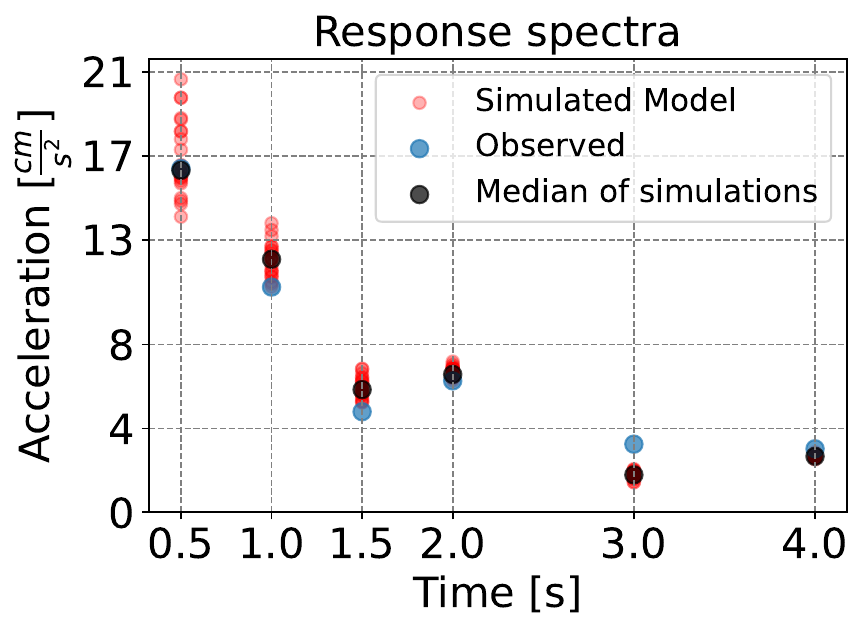}
    %\caption{Response spectra.}
   % \label{fig:third}
\end{subfigure}
\caption{Same accelerogram as in Figure~\ref{fig:UNIO6}.  Left: level curves of our fitted full model  overlapped to the ePSDF.  Right: the response spectra of 20 simulations using our fitted model, and their median, and the response spectra of the observed accelerogram.}
\label{fig:UNIO62}

\end{figure}

\subsection{Reduced model}
\label{sec:ReducedExamples}

The response spectrum analysis usually takes only the $S-$wave and ignore the $P-$wave \cite{ordaz1994bayesian},  \cite{arroyo2010strong},  \cite{vlachos2016multi}, \cite{tang2016generation}, \cite{sato2013fractal}.  To consider only the $S-$wave, the model can be simplified in a straightforward way.  To work only with the $S-$wave,  ignore the values before $l_1+\gamma$ and describe the accelerogram only with the $S-$wave parameters.
Namely 
 $$
 \Theta^*=\{\mu_2,\sigma_2^2,\pi_1,  \mu_{\mathfrak{f}_{21}},  \sigma^2_{\mathfrak{f}_{21}},   \mu_{\mathfrak{f}_{22}}, \sigma^2_{\mathfrak{f}_{22}},  \pi_2,  \rho, E_T\},
 $$
reducing the number of parameters to only 10.  This is considerably less than the 18 parameters in the $S-$wave modeling approach of \cite{vlachos2016multi}.

\begin{figure}[!ht]
\centering
\begin{subfigure}{0.45\textwidth}
    \includegraphics[width=\textwidth]{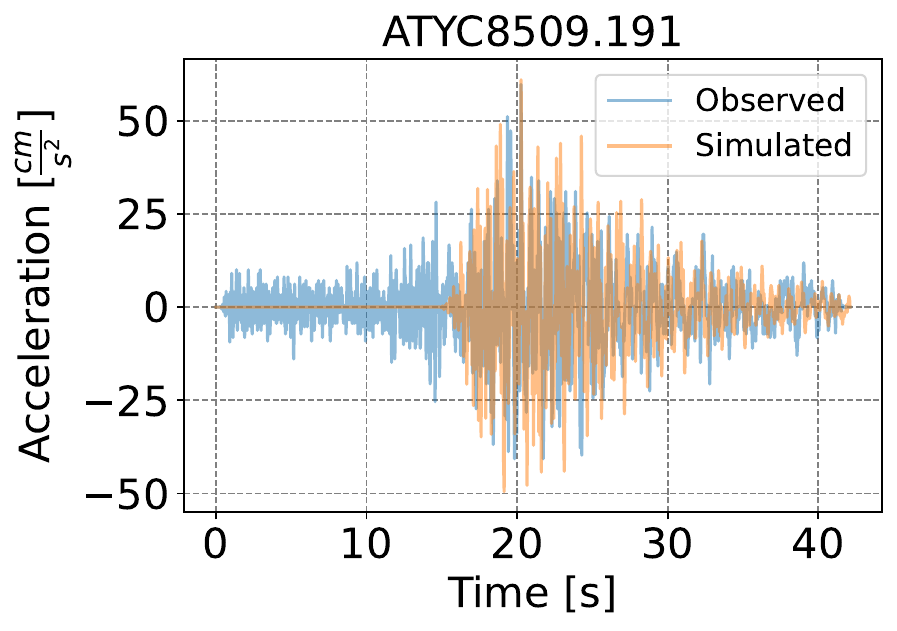}
\end{subfigure}
\begin{subfigure}{0.45\textwidth}
    \includegraphics[width=\textwidth]{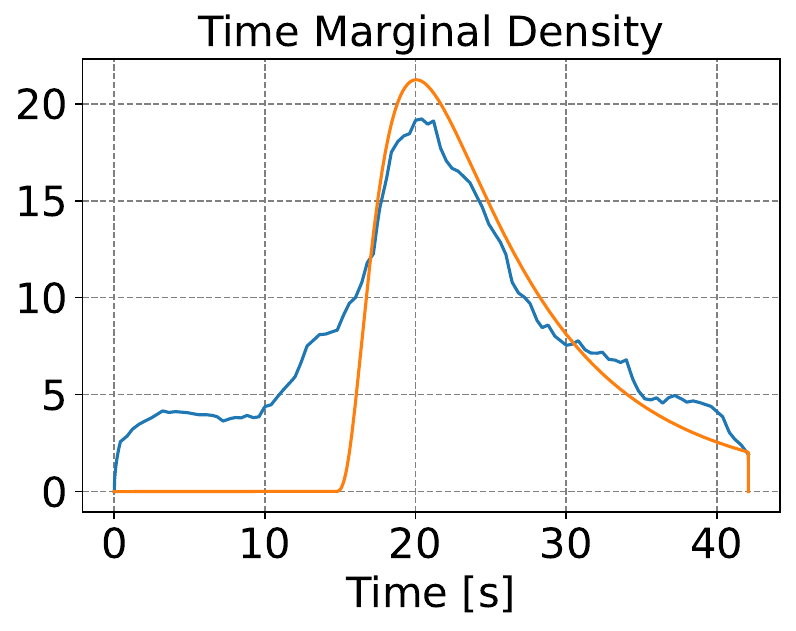}
\end{subfigure}

\begin{subfigure}{0.45\textwidth}
    \includegraphics[width=\textwidth]{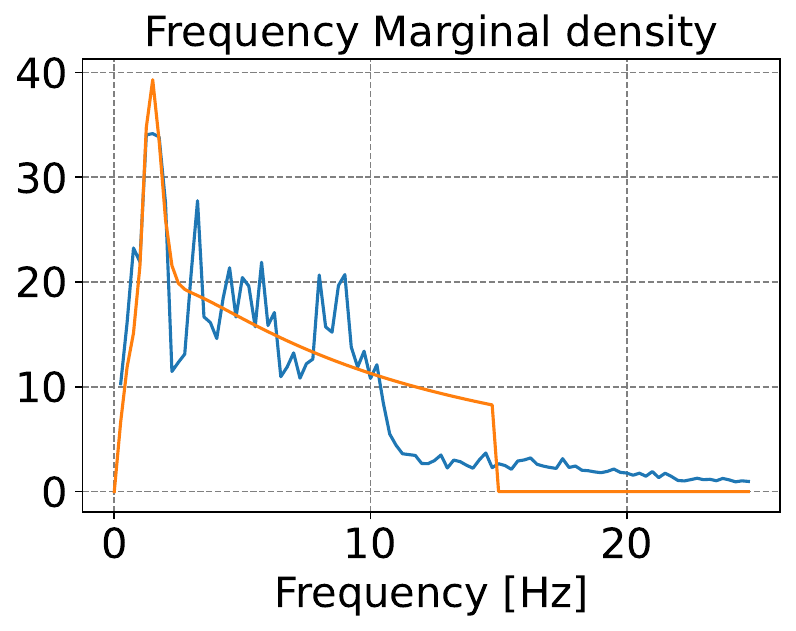}
\end{subfigure}
\begin{subfigure}{0.45\textwidth}
    \includegraphics[width=\textwidth]{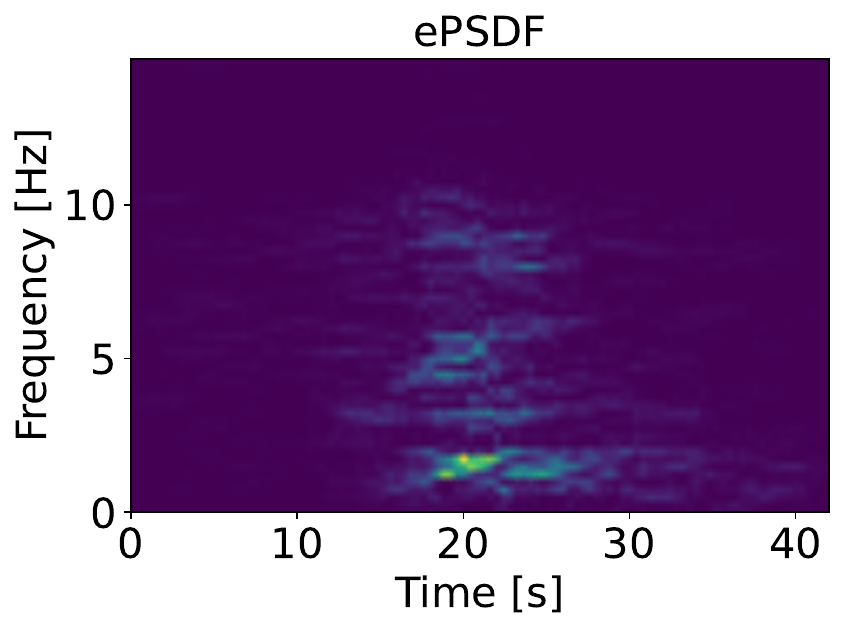}
\end{subfigure}

\caption{Summary of the fit obtained for our reduced model for the 19 September 1985 earthquake,  with $M_w=8.0$ and $R=283.09km$  recorded at Atoyac,  Guerrero.  The observed data is presented in blue,  the simulation and the fit is presented in orange.}

\label{fig:ATYCReducida}
\end{figure}

\begin{figure}[!ht]

\centering
\begin{subfigure}{0.45\textwidth}
    \includegraphics[width=\textwidth]{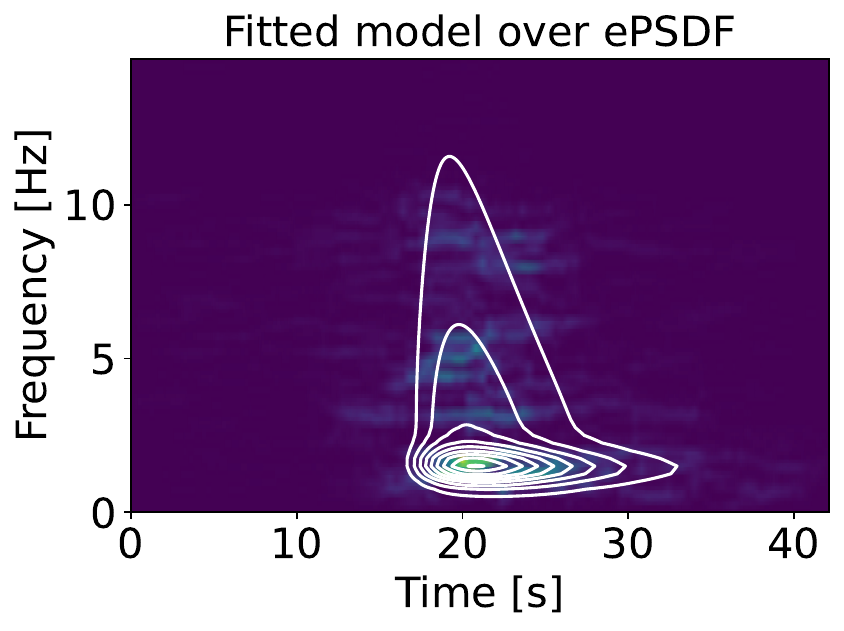}
\end{subfigure}
\begin{subfigure}{0.45\textwidth}
    \includegraphics[width=\textwidth]{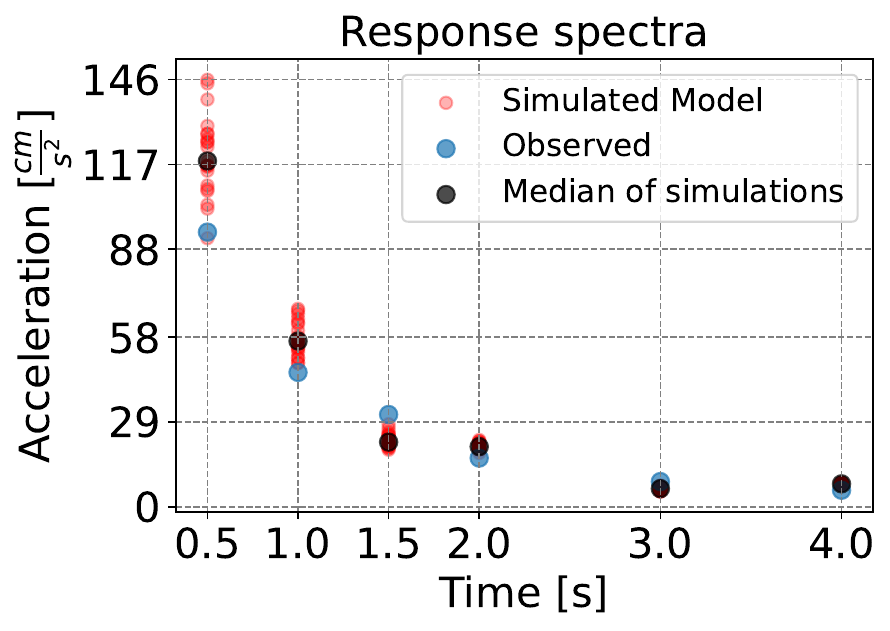}
\end{subfigure}

\caption{Left:level curves of the reduced model  overlapped to the ePSDF.  Right: Response spectra of 20 distinct simulations using the reduced model, in black its median and in blue the response spectra of the real accelerogram for the record of Figure \ref{fig:ATYCReducida}}
\label{fig:ATYCReducida2}
\end{figure}

\begin{figure}[!ht]

\centering
\begin{subfigure}{0.45\textwidth}
    \includegraphics[width=\textwidth]{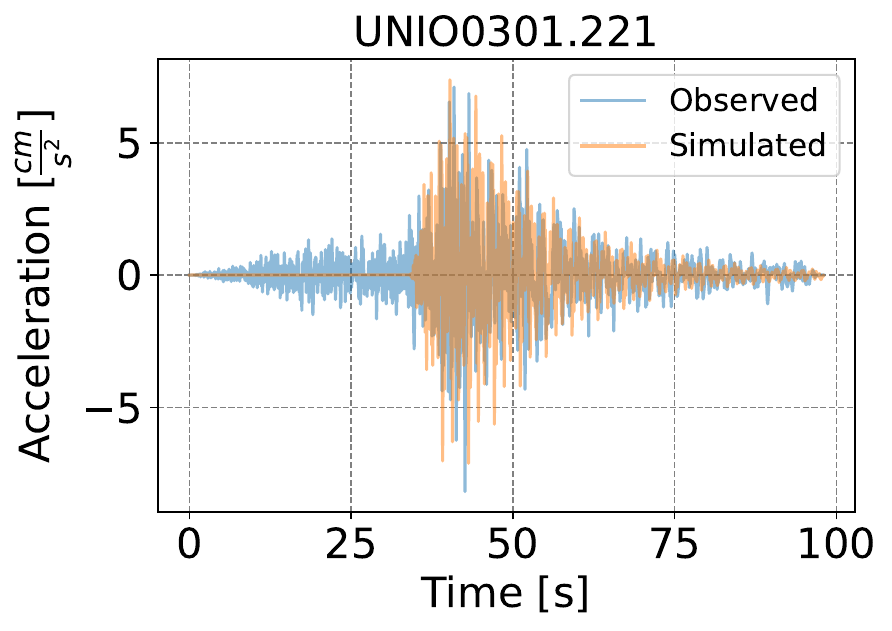}
\end{subfigure}
\begin{subfigure}{0.45\textwidth}
    \includegraphics[width=\textwidth]{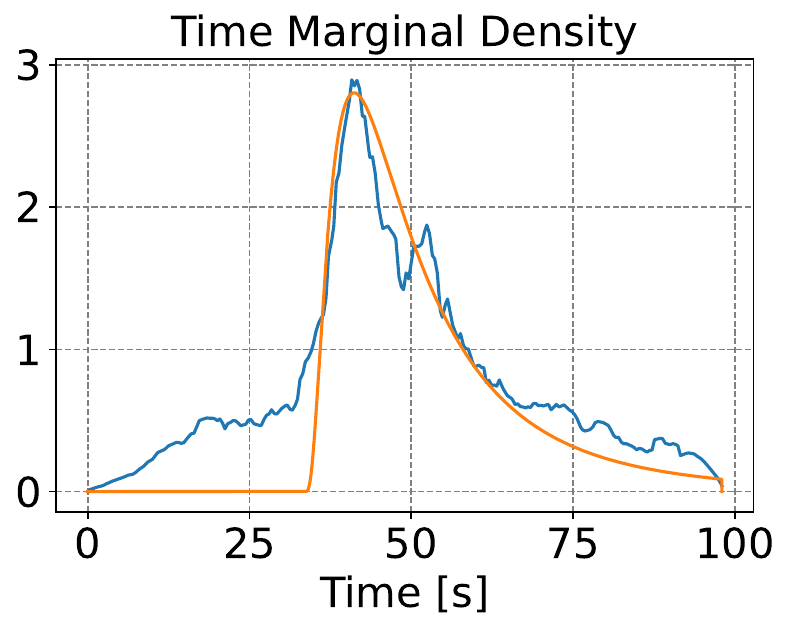}
\end{subfigure}

\begin{subfigure}{0.45\textwidth}
    \includegraphics[width=\textwidth]{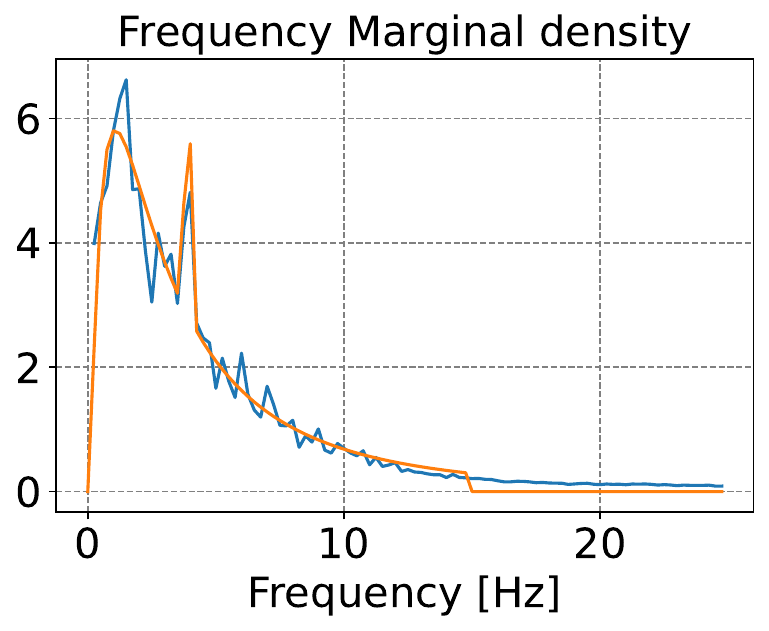}
\end{subfigure}
\begin{subfigure}{0.45\textwidth}
    \includegraphics[width=\textwidth]{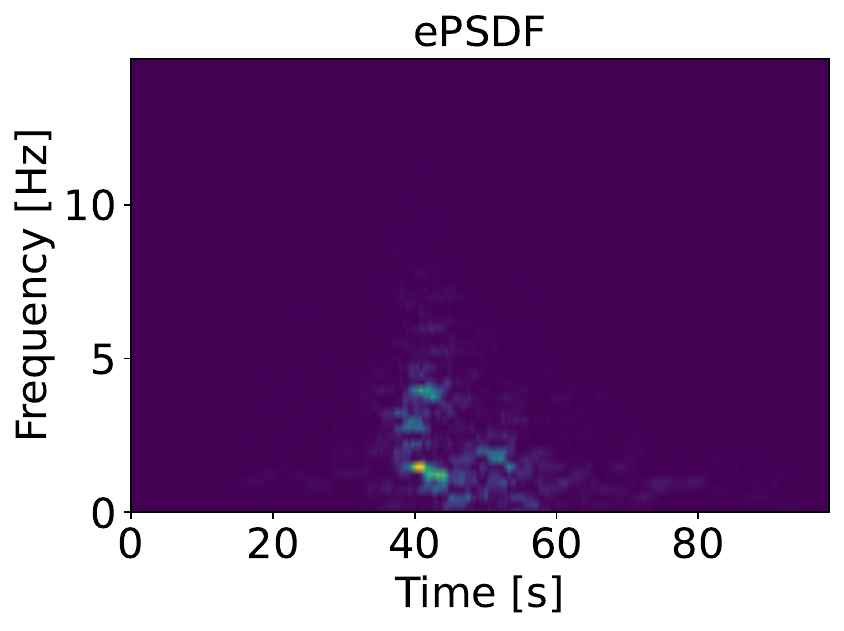}
\end{subfigure}

\caption{Summary of the fit obtained for our reduced model for the 22 January 2003 earthquake,  with $M_w=7.5$ and $R=263.59km$  recorded at La Uni\'on,  Guerrero.  The observed data is presented in blue,  the simulation and the fitting is presented in orange.}

\label{fig:UNIOReducida}
\end{figure}

\begin{figure}[!ht]
\centering
\begin{subfigure}{0.45\textwidth}
    \includegraphics[width=\textwidth]{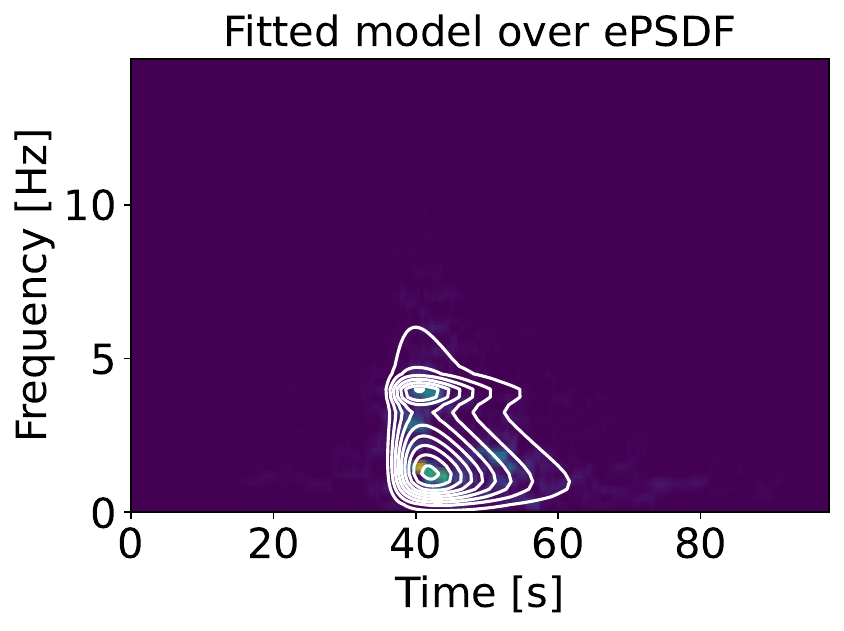}
 \end{subfigure}
\begin{subfigure}{0.45\textwidth}
    \includegraphics[width=\textwidth]{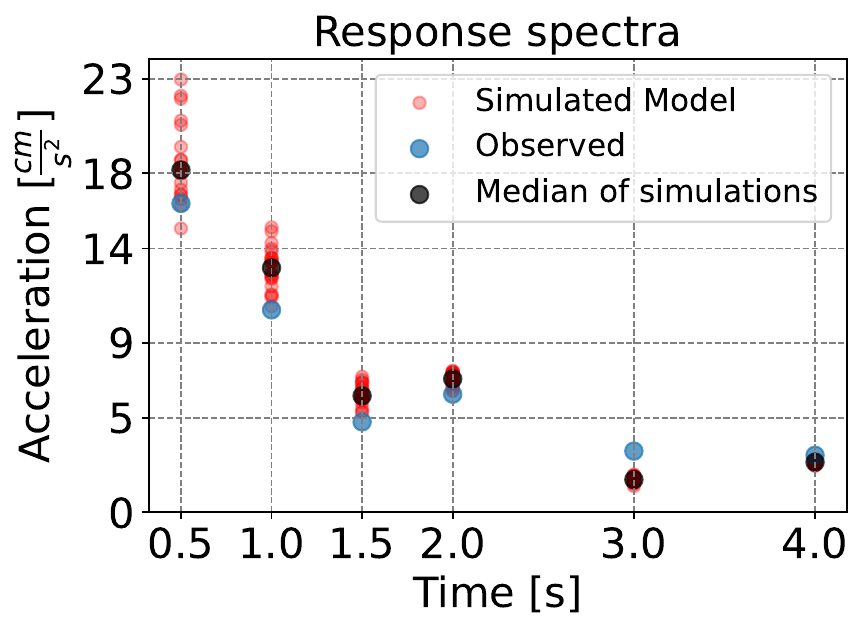}

\end{subfigure}
\caption{Left:level curves of the reduced model  overlapped to the ePSDF.  Right: Response spectra of 20 distinct simulations using the reduced model, in black its median and in blue the response spectra of the real accelerogram for the record of Figure \ref{fig:UNIOReducida}}
\label{fig:UNIOReducida2}
\end{figure}

Figures \ref{fig:ATYCReducida} and \ref{fig:UNIOReducida} present examples of the reduced model with the same accelerograms presented in Figures \ref{fig:ATYC6} and \ref{fig:UNIO6}, respectively.  It is important to note that although the reduced model is less flexible, the response spectra, in Figures \ref{fig:ATYCReducida2} and \ref{fig:UNIOReducida2}, are comparable with those obtained from the full model, in Figures \ref{fig:ATCY62} and \ref{fig:UNIO62} .

There is a trade off to take into account,  the reduced model is faster and easier to fit and allows  for more parsimonious attenuation laws,  but it describes the ePSDF less accurately, which will slightly increase the error.  Depending on the importance of the $P-$available computing power, the reduced model may be preferred over the full model.

\subsection{Error analysis} 
\label{sec:ErrorAnalysis}

To analyze the performance of the full model, the response spectra   using the synthetic accelerograms are compared to the observed ones. The procedure was as follows.  For each recorded accelerogram, $x_o$, 20 synthetic accelerograms, $x_i$ for $i=1,2,...,20$, were simulated using $E_T\hat{f}_t(\mathfrak{f})$ as their ePSDF. Subsequently, with the synthetic accelerograms, the response spectra, $\{S_a(\omega_0,0.05,x_i): i=1,2,...,20\}$, were evaluated for the natural periods of resonance $T$ in $\{0.5,1,1.5,2,3,4\}$. Finally, the median of the response spectra at time $T$ for the 20 simulations, $m(T,\{x_i: i=1,...,,20\})$, was compared with the value obtained from the observed accelerogram.

\begin{figure}[!ht]
\centering
\begin{subfigure}{0.45\textwidth}
    \includegraphics[width=\textwidth]{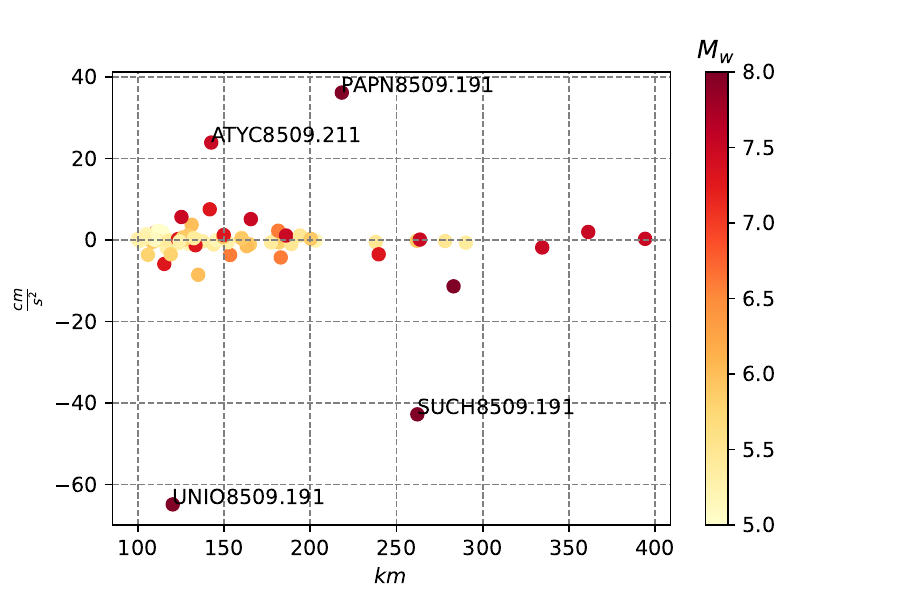}
    \caption{Period of $0.5s$.}
    %\label{fig:first}
\end{subfigure}
\begin{subfigure}{0.45\textwidth}
    \includegraphics[width=\textwidth]{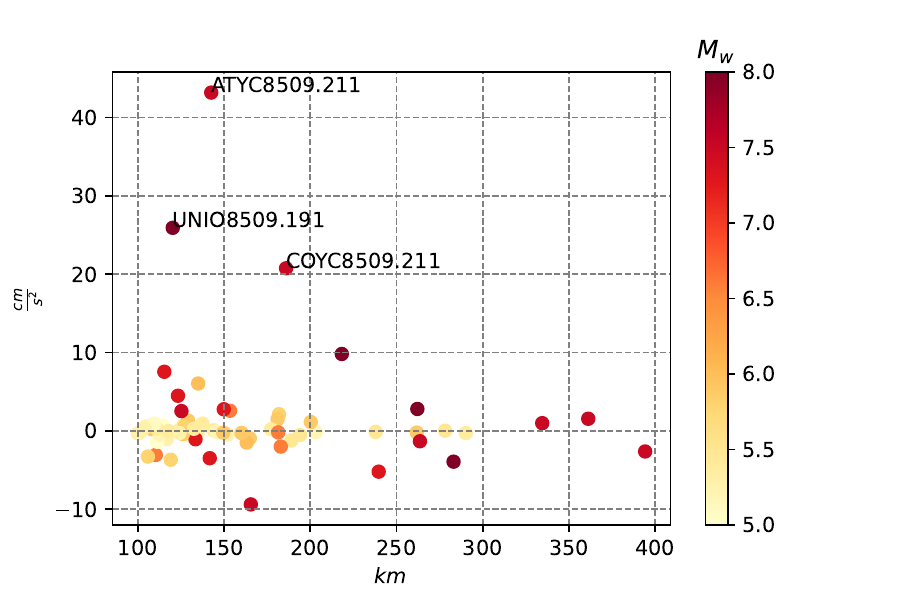}
    \caption{Period of $1s$.}
    %\label{fig:second}
\end{subfigure}

\begin{subfigure}{0.45\textwidth}
    \includegraphics[width=\textwidth]{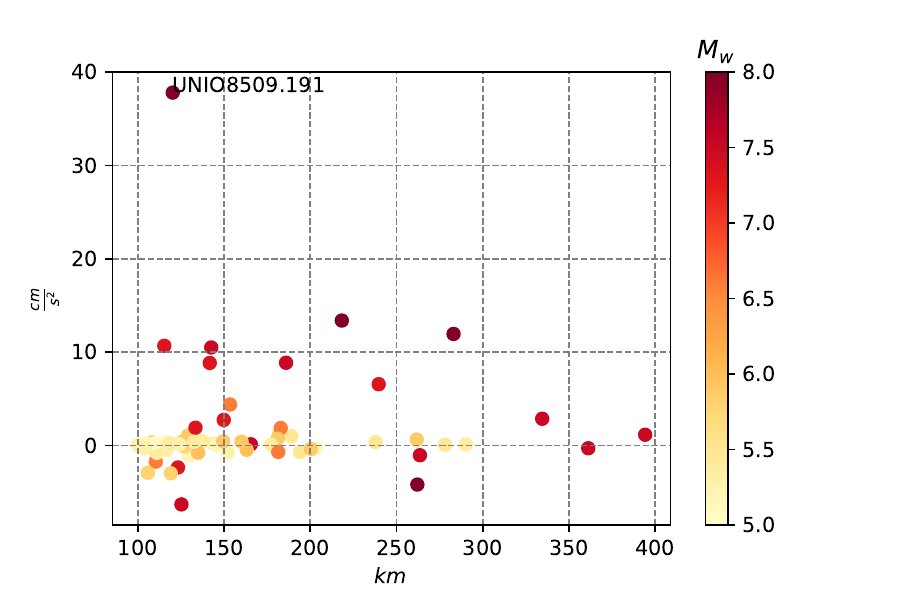}
    \caption{Period of $1.5s$.}
    %\label{fig:third}
\end{subfigure}
\begin{subfigure}{0.45\textwidth}
    \includegraphics[width=\textwidth]{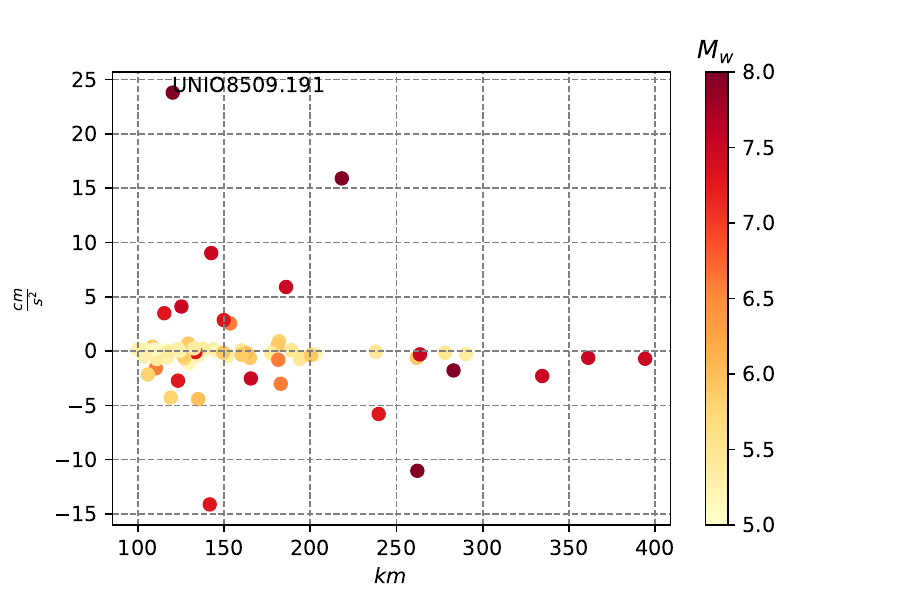}
    \caption{Period of $2s$.}
    %\label{fig:first}
\end{subfigure}

\begin{subfigure}{0.45\textwidth}
    \includegraphics[width=\textwidth]{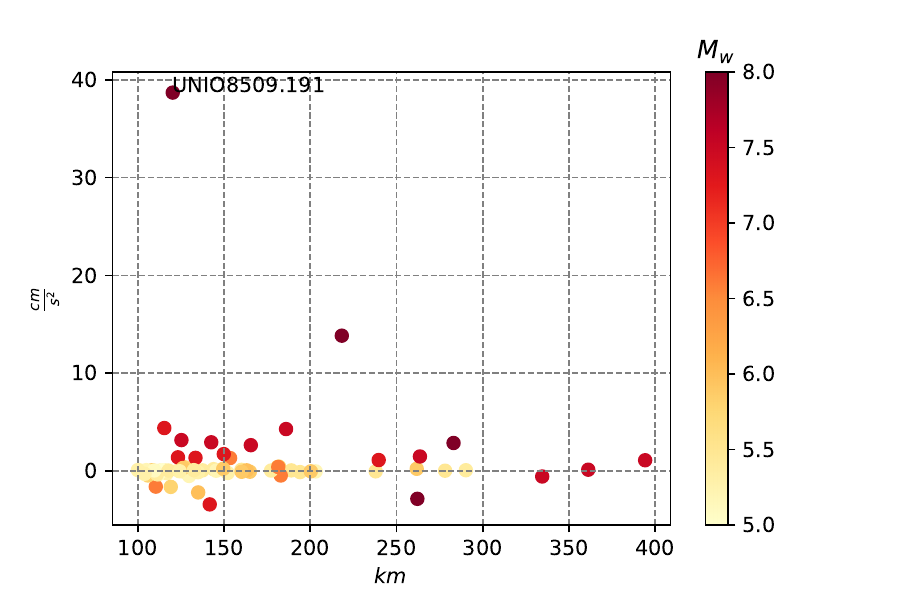}
    \caption{Period of $3s$.}
    %\label{fig:second}
\end{subfigure}
\begin{subfigure}{0.45\textwidth}
    \includegraphics[width=\textwidth]{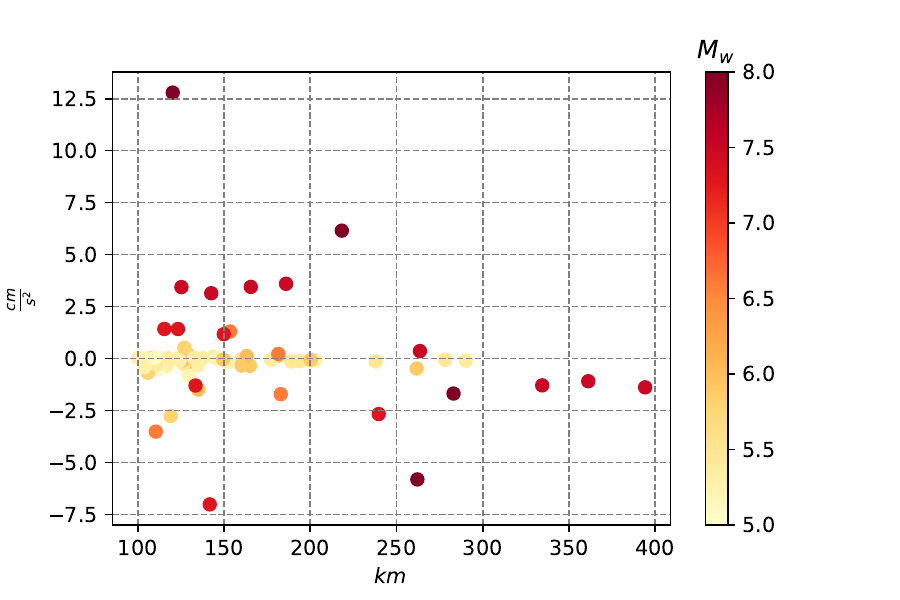}
    \caption{Period of $4s$.}
    %\label{fig:third}
\end{subfigure}
                
\caption{Difference of the response spectra of the observed accelerogram versus the median of the response spectra of 20 simulation using the full model.}
\label{fig:Error}
\end{figure}

Figure \ref{fig:Error} displays the errors of the proposed model for each record in our database. The error is simply defined as 
$$
E(\cdot,x_o)=S_a(\cdot,0.05,x_o)-m(\cdot,\{x_i: i=1,2,...,20\}).
$$ 

The errors do not have trend and almost all of them are less than $20$ gals. Moreover, the error band decreases with the period, reaching a band of $5$ gals for a period of $4$ s. These results are competitive with the errors in the response spectra shown in \cite{arroyo2010strong} and  \cite{vlachos2016multi}.

 It is important to note that the proposed model assumes a parametric representation of a non-stationary process, which has fewer parameters than similar ePSDF models, and that the inclusion of a priori physical knowledge in the marginals allows for a flexible and competitive model with parameters that are easy to interpret. 

Reviewing the errors in the figure \ref{fig:Error}, the largest error corresponds to the record UNIO8509.191 for the period of $0.5$ s with a value of $-64.90$cm/s$^2$. When the relative error is calculated, it turns out that the estimate varies only 21\% with respect to the response spectrum data. The error is large compared to other earthquakes due to the high accelerations recorded for an earthquake of $M_w=8$. In the supplementary material a good approximation of the ePSDF can be seen, although the energy release was highly scattered. Similar situations occur for the records PAPN8509.191 and SUCH8509.19 of the 19 September 1985 earthquake.

The records ATYC8509.211 and COYC8509.211 are from the earthquake of 21 september 1985 earthquake, which had a magnitude of $M_w=7.5$. In both cases, our model underestimated the frequencies below 10  Hz, resulting in a smaller predictions of the response spectra.

\section{Further work on attenuation laws}
\label{sec:Atenuaciones}

Each pair of $(\mu_i,\sigma_i^2)$ of a lognormal distribution is  information about the location and variability of the energy released,  which allows us to easily understand  the physical meaning of our parameters in $\Theta^*$, which is an important advantage over the model parameterization of \cite{vlachos2016multi} or the non-parametric model in \cite{broccardo2017spectral}. 

\begin{figure}[!ht]
\centering
\begin{subfigure}{0.45\textwidth}
    \includegraphics[width=\textwidth]{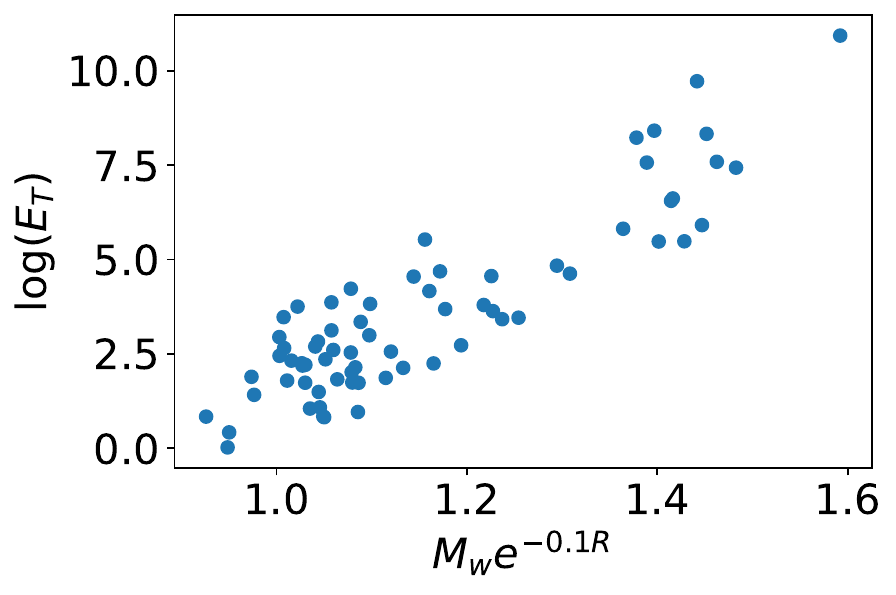}
    \caption{$(E_T, M_w e^{0.1R})$.}
    %\label{fig:first}
    \label{fig:Regresion}
\end{subfigure}
\begin{subfigure}{0.45\textwidth}
    \includegraphics[width=\textwidth]{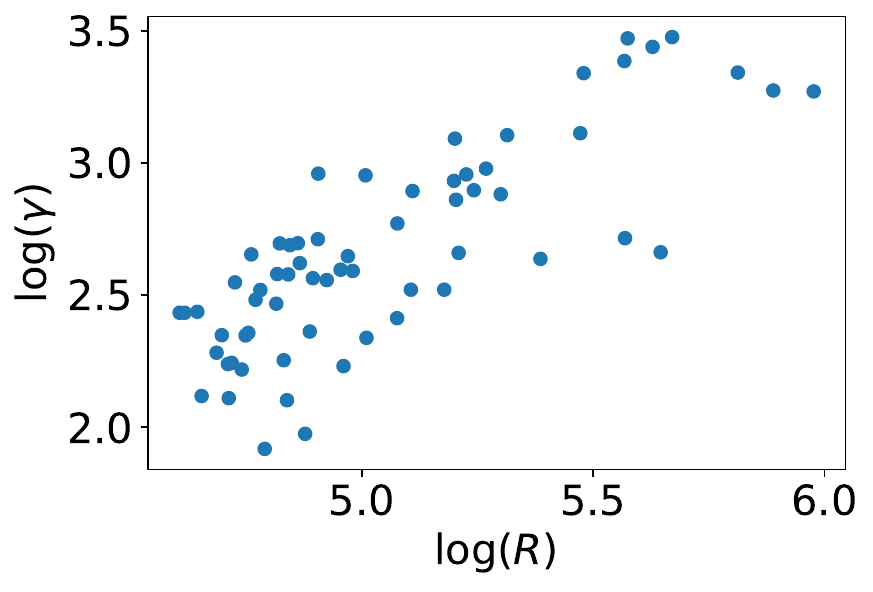}
    \caption{$(\log(R),\log(\gamma))$.}
    %\label{fig:second}
    \label{fig:Regresion2}
\end{subfigure}
\caption{The scatter plots were made using all the records.} 
\label{fig:Regresiones}
\end{figure}

To use our model in the context of attenuation laws, it will be necessary  to  predict $E_T$  from  $M_w$ and $R$.  As a first  approach  it is proposed  the heuristic relation
\begin{equation}
\label{eq:regresion}
E_T=M_w e^{0.1R}.
\end{equation}
Figure \ref{fig:Regresion} is a  scatter plot of our proposed relation, showing that  simple regression models could  predict $E_T$.

For the  lag time between waves,  Figure \ref{fig:Regresion2} is a scatter plot  of the epicenter  distance  versus  lag   time between copulas,  i.e.  the $\gamma$ parameter,  which shows a useful relationship for predicting $\gamma$  from $R$ only.

With  this evidences, it is encouraging to further extend our model for attenuation laws in the context of seismic risk, since this  novel approach for the attenuation laws reach an important complexity reduction compared with \cite{vlachos2018predictive} and it avoids the assumption of stationarity  in contrast to  the papers based in the methodology of  \cite{boore2003simulation}

%%%%%%%%%%%%%%%%%%%%%%%%%%%%%%%%%%%%%%%%%%%%%%%%%

%%%%%%%%%%%%%%%%%%%%%%%%%%%%%%%%%%%%%%%%%%%%%%%%%

\section{Conclusions}

\label{sec:Conclusion}

 The development of a fully physically inspired model that considers $P$ and $S$ waves simultaneously using only 16 scalar parameters to analyze an accelerogram was presented.   Furthermore, the amplitude spectrum or seismic envelope models could be used to generate attenuation models.  The framework presented also provides a flexible and easily interpretable model that produces errors that are competitive with other recently developed models.
 %the state of the art. THIS IS THE STATE OF THE ART!!!

The main advantage of the presented framework  is that, given the copula, the time and frequency marginals are independent, allowing each marginal to be modeled separately to finally obtain the joint distribution with a copula.

Unlike the models based on Kanai-Tajimi filters, which require an additional high-pass filter to reduce the energy content in the lower frequencies, the model presented in  this paper allows the energy decay for the lower frequencies to be controlled by the probability density function.

Even though the knowledge of the marginals in our model is designed for the specific region of the Southern Mexico, the model is flexible enough to  be easily modified to  work with  other regions.

Implementing these ideas for ground motions with $R>500$, the addition of surface waves could easily be done by adding more elements to the mixtures of the marginals or using probability density functions with light tails for the frequency margin to concentrate the energy in a smaller range of frequencies.

Although our model uses the $P$ and $S$ wave  arrival times  to constrain the optimization problem, if  they are known, they  could be fixed for a better inference. It is also important to note that the full model does not try to identify each wave with high accuracy, our model only uses the information of the waves to create an adequate envelope of the earthquake.

All the codes presented in this work can be found in the repository \url{https://github.com/isaiasmanuel/Attenuation_Laws} and are available for download.

%%%%%%%%%%%%%%%%%%%%%%%%%%%%%%%%%%%%%%%%%%%%%%%%%

%%%%%%%%%%%%%%%%%%%%%%%%%%%%%%%%%%%%%%%%%%%%%%%%%

\appendix

\section{Basics of Copulas}
\label{sec:BasicsCopulas}

In order to have a self-contained text, the definition of a copula and the Sklar's theorem are presented.

\begin{definition}
A two-dimensional copula is a function $C$ with the following properties:
\begin{enumerate}
\item $C$ domain is $[0,1]\times [0,1]$
\item $C(x,0)=0=C(0,y)$ for all $(x,y)$ in $[0,1]\times [0,1]$ 
\item If $0\leq x_1\leq x_2 \leq 1$ and $0\leq y_1\leq y_2\leq 1$ then
$$
C(x_2,y_2)-C(x_2,y_1)-C(x_1,y_2)+C(x_1,y_1)\geq 0
$$
\item For every $(u,v)$ in $[0,1]\times [0,1]$ is met
$$
C(u,1)=u\quad\text{and}\quad C(1,v)=v.
$$
\end{enumerate}
\end{definition}

A copula is usually interpreted as a bivariate distribution function on the unitary square with uniforms in $[0,1]$ as marginal distributions.

A key theorem in copulas theory is Sklar's theorem, which relates every bivariate distribution function to an associated  copula; moreover, to define a bivariate continuous distribution, it is only necessary to know the marginal distributions and the  associated copula.

\begin{theorem}{Sklar's Theorem.}
\label{SklarTheorem}
Let $H$ be a joint distribution function with margins $F$ and $G$. Then there exists a copula $C$ such that for all $x,y$ in $\overline{\mathbb{R}}$,
\begin{equation}
\label{eq:sklar}
H(x,y)=C(F(x),G(y)).
\end{equation}
\end{theorem}

If $F$ and $G$ are continuous,  then $C$ is unique; otherwise, $C$ is uniquely determined on the cross product  $\text{rank}(F) \times \text{rank}(G)$.  Conversely,  if $C$ is a copula and $F$ and $G$ are distribution functions, then the function $H$ defined by \eqref{eq:sklar} is a joint distribution function with marginal distributions $F$ and $G$.

The corollary \ref{Recupera} gives a direct way to recover the copula associated to a joint distribution,  it is very useful to define copulas with properties associated to a specific distribution.

\begin{corollary}
\label{Recupera}
Let $H,F,G$ and $C$ as in Theorem \ref{SklarTheorem} and let $F^{(-1)}$ and $G^{(-1)}$ be the quantile functions of $F$ and $G$,  respectively. Then for any $(u,v)$ in $[0,1]\times [0,1]$,
$$
C(u,v)=H(F^{-1}(u),G^{-1}(v)). 
$$
\end{corollary}

\section{Response spectrum}
\label{sec:RVT}

A very important ground motion parameter is the response spectrum, which is discussed in detail in \cite{kramer1996geotechnical}.   A main idea during the development of the  model  was  to describe seismic motion  that can reproduce  the response spectrum.  In this section, the response spectrum is briefly introduced. 

For an angular frequency $\omega_0$ and a damping ratio $\zeta$, the displacement $u(t)$ is modeled by a forced damped harmonic oscillator, i.e. it follows
\begin{equation}
\label{eq:oscillator}
\frac{d^2u}{dt^2}+2\zeta\omega_0\frac{du}{dt}+\omega_0^2 u=-x(t),
\end{equation}
with initial conditions
\begin{align*}
u(0)=0\\
\frac{d}{dt}u(0)=0,
\end{align*}
where $x$ is an accelerogram.  

The response acceleration spectrum of $x$ is $\max_t  \mid \frac{d^2u}{dt^2} \mid  .$

As it is discussed in \cite{kramer1996geotechnical} is common to take the pseudo acceleration response spectrum (PARS), defined by 

\begin{equation}
\label{eq:PseudoresponseSpectra}
S_a(\omega_0,\zeta, x)= \omega_0^2 \max_t  \mid  u \mid ,
\end{equation}
which is an approximation of the response acceleration spectrum.  In this work the response spectra were computed with equation \eqref{eq:PseudoresponseSpectra}.

To solve the equation \eqref{eq:oscillator} the SciPy extension LSODA on Python was used, in where Adams and BDF method with automatic stiffness detection and switching are implemented.  The damping parameter was set to $\xi=0.05$, as it is the common value used in the  specialized  literature,  as in \cite{arroyo2010strong},  \cite{boore2003simulation},  \cite{bora2016relationship},  \cite{barone2015novel}, \cite{alimoradi2015machine},  among others.

\printcredits

\bibliographystyle{cas-model2-names}

\bibliography{cas-refs}


%Author biography without author photo.

%\bio{figs/pic1}
%Author biography with author photo.

%\bio{figs/pic1}
%Author biography with author photo.

\end{document}